\documentclass[a4paper,UKenglish,cleveref, autoref, thm-restate]{lipics-v2021}

\usepackage{algorithm, algorithmic}
\usepackage{tikz}

\newcommand*{\0}{\phantom{0}}
\newcommand{\crule}[1]{\multispan{#1}{\hspace*{\tabcolsep}\hrulefill
  \hspace*{\tabcolsep}}}
\pdfoutput=1
\hideLIPIcs

\title{Multi-qubit lattice surgery scheduling}

\author{Allyson Silva}{1QB Information Technologies (1QBit), Vancouver, Canada}{allyson.silva@1qbit.com}{https://orcid.org/0000-0001-6807-0548}{}

\author{Xiangyi Zhang}{1QB Information Technologies (1QBit), Vancouver, Canada}{xiangyi.zhang@1qbit.com}{https://orcid.org/0000-0003-4395-2776}{}

\author{Zak Webb}{1QB Information Technologies (1QBit), Vancouver, Canada}{zachary.webb@1qbit.com}{https://orcid.org/0000-0003-1071-5467}{}

\author{Mia Kramer}{1QB Information Technologies (1QBit), Vancouver, Canada}{mia.kramer@1qbit.com}{https://orcid.org/0009-0001-6962-1863}{}

\author{Chan W. Yang}{1QB Information Technologies (1QBit), Vancouver, Canada}{chanwoo.yang@1qbit.com}{https://orcid.org/0009-0003-0556-8981}{}

\author{Xiao Liu}{1QB Information Technologies (1QBit), Vancouver, Canada}{}{https://orcid.org/0009-0000-9687-7293}{}

\author{Jessica Lemieux}{1QB Information Technologies (1QBit), Vancouver, Canada}{}{https://orcid.org/0000-0001-6884-0282}{}

\author{{Ka-Wai} Chen}{1QB Information Technologies (1QBit), Vancouver, Canada}{}{}{}

\author{Artur Scherer}{1QB Information Technologies (1QBit), Vancouver, Canada}{artur.scherer@1qbit.com}{https://orcid.org/0000-0001-5247-656X}{}

\author{Pooya Ronagh\footnote{Corresponding author}}{1QB Information Technologies (1QBit), Vancouver, Canada \and Institute for Quantum Computing, University of Waterloo, Waterloo, Canada \and Department of Physics \& Astronomy, University of Waterloo, Waterloo, Canada \and Perimeter Institute for Theoretical Physics, Waterloo, Canada}{pooya.ronagh@1qbit.com}{https://orcid.org/0000-0002-9591-9727}{Acknowledges the financial support of Mike and Ophelia
Lazaridis, Innovation, Science and Economic Development
Canada (ISED), and the Perimeter Institute for
Theoretical Physics. Research at the Perimeter Institute
is supported in part by the Government of Canada
through ISED and by the Province of Ontario through
the Ministry of Colleges and Universities. }

\authorrunning{A. Silva et al.}

\Copyright{Allyson Silva, Xiangyi Zhang, Zak Webb, Mia Kramer, Chan W. Yang, Xiao Liu, Jessica Lemieux, {Ka-Wai} Chen, Artur Scherer, and Pooya Ronagh}

\ccsdesc[500]{Computer systems organization~Quantum computing}
\ccsdesc[300]{Software and its engineering~Compilers}
\ccsdesc[300]{Hardware~Quantum error correction and fault tolerance}
\ccsdesc[300]{Software and its engineering~Scheduling}
\ccsdesc[300]{Hardware~Circuit optimization}

\keywords{Scheduling, Large-Scale Optimization, Surface Code, Quantum Compilation, Circuit Optimization}

\relatedversion{}
\relatedversiondetails{Full Version}{https://arxiv.org/abs/2405.17688}

\supplement{}
\supplementdetails[subcategory={Quantum circuits used for the experiments}, cite={}, swhid={}]{Dataset}{https://doi.org/10.5281/zenodo.11391890}

\acknowledgements{We are grateful to our editor, Marko Bucyk, for his careful review and editing of the manuscript.}

\nolinenumbers

\begin{document}

\maketitle

\begin{abstract}
Fault-tolerant quantum computation using two-dimensional topological quantum
error correcting codes can benefit from multi-qubit long-range operations. By
using simple commutation rules, a quantum circuit can be transpiled into a
sequence of solely non-Clifford multi-qubit gates. Prior work on fault-tolerant
compilation avoids optimal scheduling of such gates since they reduce the
parallelizability of the circuit. We observe that the reduced parallelization
potential is outweighed by the significant reduction in the number of gates. We
therefore devise a method for scheduling multi-qubit lattice surgery using an
earliest-available-first policy, solving the associated forest packing problem
using a representation of the multi-qubit gates as Steiner trees. Our extensive
testing on random and various Hamiltonian simulation circuits demonstrates the
method's scalability and performance. We show that the transpilation
significantly reduces the circuit length on the set of circuits tested, and that
the resulting circuit of multi-qubit gates has a further reduction in the
expected circuit execution time compared to serial execution.
\end{abstract}

\section{Introduction}
\label{sec:introduction}

Fault-tolerant quantum computation (FTQC) aims to ensure reliable quantum
computing despite faulty physical qubits. In FTQC, quantum error correction
(QEC) is used to protect a logical Hilbert space within a much larger one.
Topogical quantum error correcting codes in two dimensions, such as surface
codes \cite{fowler2012surface}, are of particular interest given the
convenience of nearest neighbour interactions for physical realization of
quantum computers. FTQC can be achieved on topological error correction codes
using \textit{lattice surgery}, which facilitates long-range entanglement via
auxiliary topological patches~\cite{horsman2012surface}.

At the physical level, the circuits executed during FTQC, are repeated rounds
of parity check operations that are scheduled to perform desired logical
gates. The logical qubits (codes) that are not involved in a logical gate
must still be protected using rounds of parity checks. Therefore, minimizing
the depth of the logical circuit by parallelizing the gates reduces the total
accumulated error during computation. Our work addresses the problem of
scheduling these quantum operations on a fault-tolerant architecture using
lattice surgery, which we refer to as the \textit{lattice surgery scheduling
problem} (LSSP). Efficient methods for solving the LSSP not only serve as
foundations for future quantum compilers but are also immediately applicable for
predicting the quantum resources required for target quantum algorithms.
We will focus on surface codes in the rest of this paper; however, our approach
to the LSSP is easily generalizable to other two-dimensional topological codes.

A fault-tolerant algorithm can be represented as a
sequence of Clifford and non-Clifford Pauli rotations \cite{litinski2018game}. The Clifford gates
are commuted to the end of the circuit and past the logical measurements,
resulting in a solely non-Clifford sequence of logical gates. We call this
step transpilation. This procedure is perceived to have two drawbacks: (1) it is
computationally expensive to iteratively apply a set of commutation rules to
pairwise consecutive gates to achieve this circuit, and (2) the resulting
non-Clifford gates are highly non-local and therefore less parallelizable. The
latter caveat motivates \cite{beverland2022surface, watkins2024high} to avoid
this transpilation and use algorithms for solving various shortest path
problems to parallelize circuits involving single- and two-qubit gates.

The first drawback can be rectified using a result known to the community
(explicitly explained in Appendix D of \cite{kim2022fault-tolerant}) using the
symplectic representation of Clifford gates to implement an efficient
transpiler. By sweeping over the entire circuit, a
symplectic representation is updated through commutation events. This procedure
scales linearly with the total number of logical gates (as opposed to quadratic
scaling of na\"ive usage of pairwise commutation rules). Then commuting layers
of non-Clifford gates are formed, and used to combine some of these gates into
Clifford ones. The procedure of commuting Clifford operations out is then
repeated on the new sequence, until convergence is achieved.

As for the second perceived drawback, we show in \cref{sec:results} that
the reduction in gate count from the transpilation process greatly exceeds the
reduction in parallelizability of realistic circuits. We also observe that
there is still a significant parallelizability potential between the
highly non-local resultant gates, motivating us to solve the LSSP by devising
greedy heuristics for solving the NP-hard terminal Steiner tree
problem~\cite{lin2002terminal}. We do so by decomposing the LSSP into forest
packing problems, another variant in the Steiner tree problem family~\cite
{lau2005packing, gassner2010steiner}. This entails generating Steiner trees
that connect the qubits required by each parallelized operation without
involving overlapping resources.

Our scheduling algorithm's performance is evaluated on a diverse set of
circuits, including some generated to simulate real quantum systems, such as in
the field of quantum chemistry, with up to nearly 23 million gates prior to
optimization. We analyze the scalability and performance of our proposed
algorithms to reduce gate count and to schedule operations. Our algorithm also
allows us to compare the performance of various layouts of arrays of logical
qubits surrounded by bus qubits (see \cref{sec:surface_code} for more
details). As a corollary, we propose a layout that provides a good balance for
the space--time cost trade-off against previously suggested layouts. We
draw the following conclusions from our study:
\begin{itemize}
\item The transpilation algorithm \cite{litinski2018game} for gate
 count reduction will be essential for enabling large-scale FTQC as it reduces
 circuit length by around one order of magnitude for the circuits tested.
\item Despite the transpilation reducing parallelizability of operations, the
 resulting circuits do not require prohibitive runtimes, unlike as stated
 in~\cite{beverland2022surface}. Across all tested circuits, lower bounds
 calculated for the optimal number of logical cycles required to run
 pre-transpiled circuits are between about two and 12 times higher than upper
 bounds found for post-transpiled ones.
\item Our proposed algorithm can schedule the multi-qubit gates at a rate of
 tens of thousands of operations per second in the computational environment
 tested, meaning that large quantum circuits can be scheduled in between a few
 minutes and a couple of hours using the proposed method.
\item The parallel scheduling also results in solutions that are better than
 those of serial scheduling, with some circuits among those tested having as
 many as a third of their operations benefiting from parallelization, while
 others have as few as 0.1\%.
\end{itemize}
This paper is organized as follows. \cref{sec:surface_code} presents the surface code layouts studied, which are necessary for understanding the scheduling problem we solve.
In \cref{sec:problem}, we mathematically define the LSSP using a decomposition method which guided us in designing our heuristics. \cref{sec:method} describes the algorithms proposed to generate dependency constraints and to solve the LSSP. In \cref{sec:results}, we present the results and an analysis of our computational experiments and assess the performance of the proposed algorithms for a variety of circuits. We conclude the paper in \cref{sec:conclusion} with some remarks on our research.

\section{The surface code layout} \label{sec:surface_code}

\begin{figure}
\centering
\includegraphics[scale=0.2]{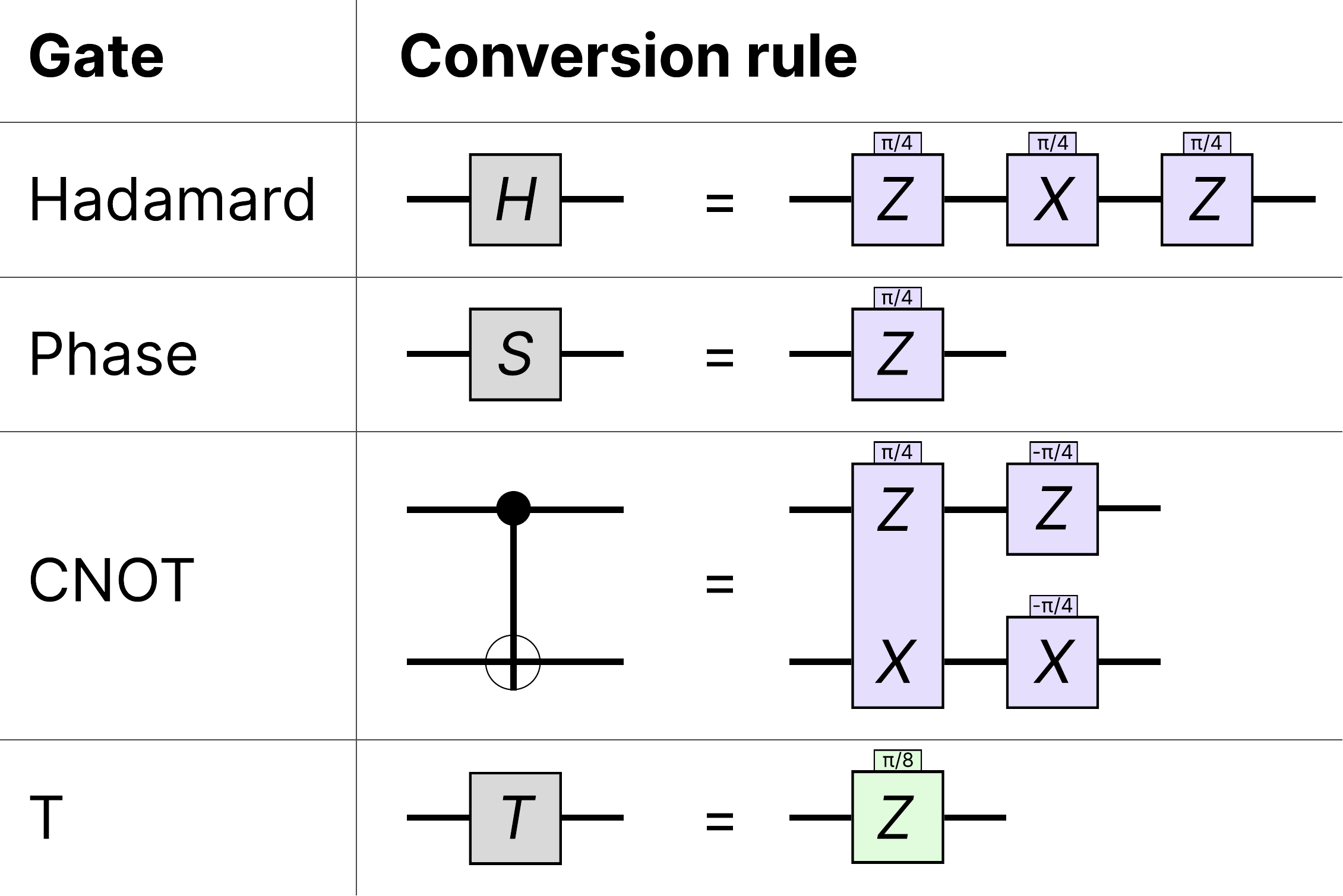}
\caption{Rules for converting Clifford + $T$ gates into Pauli product rotations. The letters within the boxes on the right ($X$, $Z$) represent the Pauli matrix, while the box colour represents the rotation angle, either $\pi/4$ (purple) or $\pi/8$ (green).}
\label{fig:basis_conversion}
\end{figure}

Following \cite{litinski2018game}, a circuit described in the Clifford + $T$ gate set consisting of the Pauli gates ($X$, $Y$, $Z$), Hadamard gates ($H$), phase gates ($S$), controlled-NOT gates (CNOT), and $T$ gates is first converted to a sequence of $\pi/4$ (Clifford) and $\pi/8$ (non-Clifford) Pauli rotations, represented as $P_\theta := \mathrm{exp}(i \theta P)$, where $P$ is a Pauli operator and $\theta$ is a rotation angle (\cref{fig:basis_conversion}).
The next step is a procedure called transpilation, in which the Clifford operations are moved through the circuit using commutation rules and eventually removed from the circuit, leaving only $\pi/8$ rotations. This process generally makes the operations less parallelizable, but also reduces the total number of rotations in the circuit. The na\"ive method for performing this transpilation in \cite{litinski2018game} takes $\mathcal{O}(m^2)$ time, where $m$ is the length of the circuit, as each Clifford operation needs to be commuted through each $\pi/8$ rotation. However, there is a faster algorithm \cite{kim2022fault-tolerant} that employs techniques similar to efficient simulations of Clifford operations that reduces this runtime to $\mathcal{O}(m)$, which for the sake of completeness is described in \cref{app:improved_gate_optimization}.

After transpilation, all Clifford rotations are removed from the circuit. In our study, the input for the LSSP is a circuit composed of $\pi/8$ rotations and the final qubit measurements. Optionally, $\pi/4$ rotations are also accepted for the scheduling of non-transpiled circuits containing Clifford gates. In our experiments described in \cref{sec:gate_opt_analysis}, we generate schedules for circuits both before and after the transpilation described in \cref{app:improved_gate_optimization}, and analyze the challenges and benefits of using this optimization procedure prior to scheduling.

\begin{figure}
\centering
\includegraphics[scale=0.5]{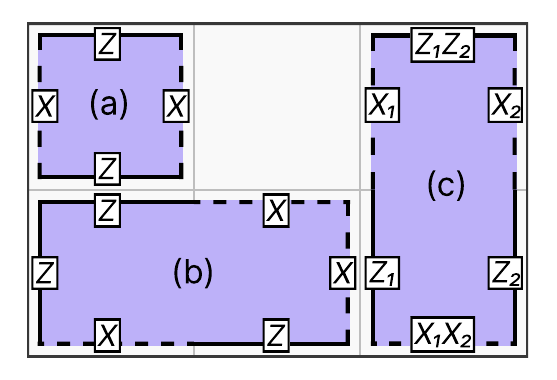}
\caption{Three types of surface code patches within a 3 $\times$ 2 grid of tiles. The edges of the patches represent the Pauli operators $X$ (dashed) and $Z$ (solid). Shown are example (a) single-tile single-qubit, (b) two-tile single-qubit, and (c) two-tile two-qubit patches. Single-qubit patches follow an $XZXZ$ pattern initialized in any position desired, such as (a) and (b). Patches can be extended to multiple tiles using lattice surgery.}
\label{fig:patch_examples}
\end{figure}

We consider a large array of physical qubits partitioned into patches of surface codes of a desired distance (see~\cref{fig:patch_examples}). Two-qubit patches provide a surface code layout where both qubits can have both $X$ and $Z$ operators accessible from each side of the patch~\cite{litinski2018game}. This way, $Y$ operators can be performed in a single step by connecting the ancilla patch to both $X$ and $Z$ operators simultaneously. Operations like patch initialization, patch measurement, and patch deformation can manipulate qubits associated with these patches~\cite{litinski2018game, watkins2024high}. Lattice surgery using ancilla patches enable long-range entangling gates between the logical qubits. We call the set of tiles dedicated to ancilla patch generation the \textit{quantum bus}, with each tile in the quantum bus hosting a \textit{bus qubit}. The qubits associated with those required by the quantum operations are called \textit{data qubits}. Ancilla patches, generated during measurements, can be discarded afterwards, freeing up bus qubits for reuse by newer ancilla patches generated for another operation.

\begin{figure}
\centering
\includegraphics[scale=0.3]{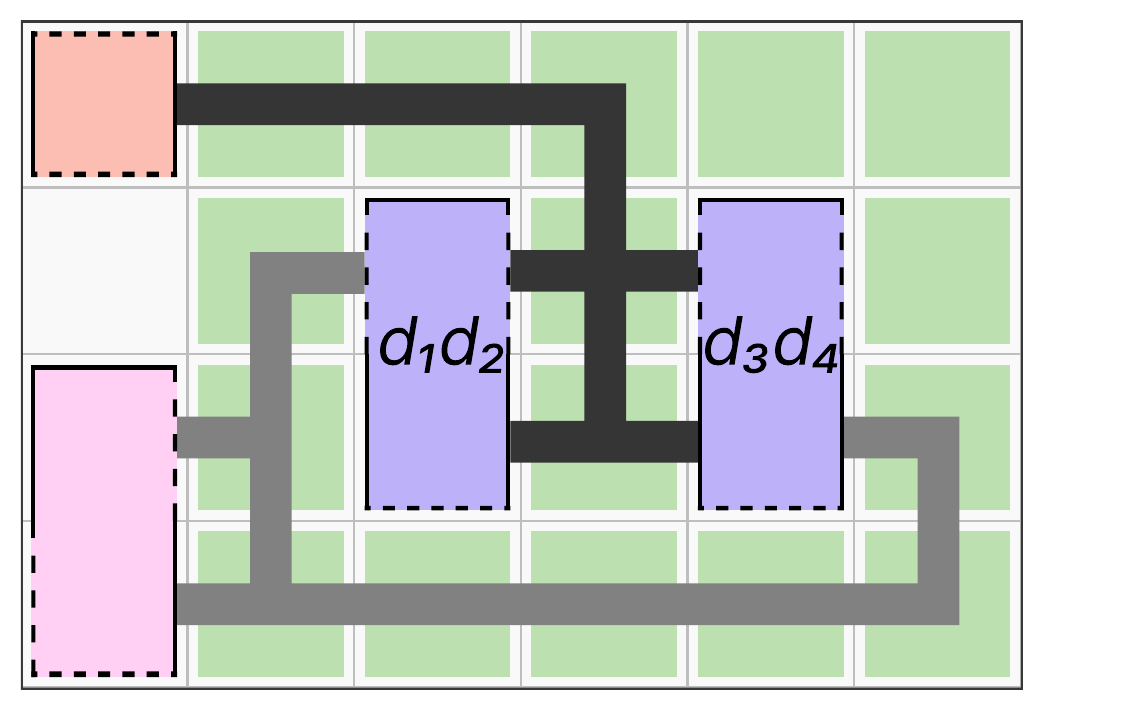}
\caption{Example of two parallel multi-qubit measurements performed using lattice surgery in a surface code grid with data qubits (shown in purple), bus qubits (green), a magic state storage qubit (red), and an ancillary qubit (pink). The $\pi/4$ rotation corresponds to an $X \otimes I \otimes I \otimes Z$ rotation connected to the ancillary qubit available using eight bus qubits, and the $\pi/8$ rotation corresponds to an $I \otimes Y \otimes Y \otimes I$ rotation connected to the magic state storage qubit available using five bus qubits.}
\label{fig:ancilla_patch}
\end{figure}

Ancilla patches can be generated in parallel to perform multiple multi-qubit measurements simultaneously, as long as they do not share a bus or a data qubit\footnote{It is unclear to us whether, with standard lattice surgery operations, a data qubit can contribute to one measurement with an $X$ or $Z$ operator and a second commuting measurement with the other operator simultaneously. Nevertheless, in \cref{sec:scalability_analysis} we show that scheduling solutions can be improved by about 2\% in randomly generated circuits if this case is allowed compared to when it is forbidden.}. Quantum operations involving $\pi/4$ and $\pi/8$ rotations require the entanglement of the qubits required to an extra qubit in a special state~\cite{bravyi2005universal}. An operation $P_{\pi/4}$ corresponds to a $P \otimes Y$ operation involving an \textit{ancillary qubit} in the zero state. Meanwhile, an operation $P_{\pi/8}$ corresponds to a $P \otimes Z$ operation involving a qubit in a special state called a \textit{magic state}. \cref{fig:ancilla_patch} shows an example of two quantum operations---a $\pi/4$ and a $\pi/8$ rotation---performed in parallel. Operations involving $\pi/8$ rotations may still require an additional corrective $\pi/2$ rotation operation for all qubits originally measured with a probability of 50\%, but this correction would take no logical cycles as they are tracked classically~\cite{litinski2018game}.

While zero states can be instantly initialized in an ancillary qubit, magic states are prepared through \textit{magic state distillation}~\cite{bravyi2012magic, litinski2019magic} which is a costly procedure. Therefore it is customary to assume that this procedure is performed in a separate dedicated zone that interacts with the area comprising the data qubits on which the logical operations are performed. We call this area the \textit{central zone}, which is connected to \textit{magic state storage qubits} located at the boundaries of this zone. Having enough magic state factories providing a distillation rate high enough to meet the magic state consumption rate within the central zone guarantees a continuous supply of magic states to the magic state storage qubits with no overhead required to be taken into account in the LSSP.

\begin{figure}[t]
\centering
\begin{tabular}{ccccccc}
    \includegraphics[scale=0.15]{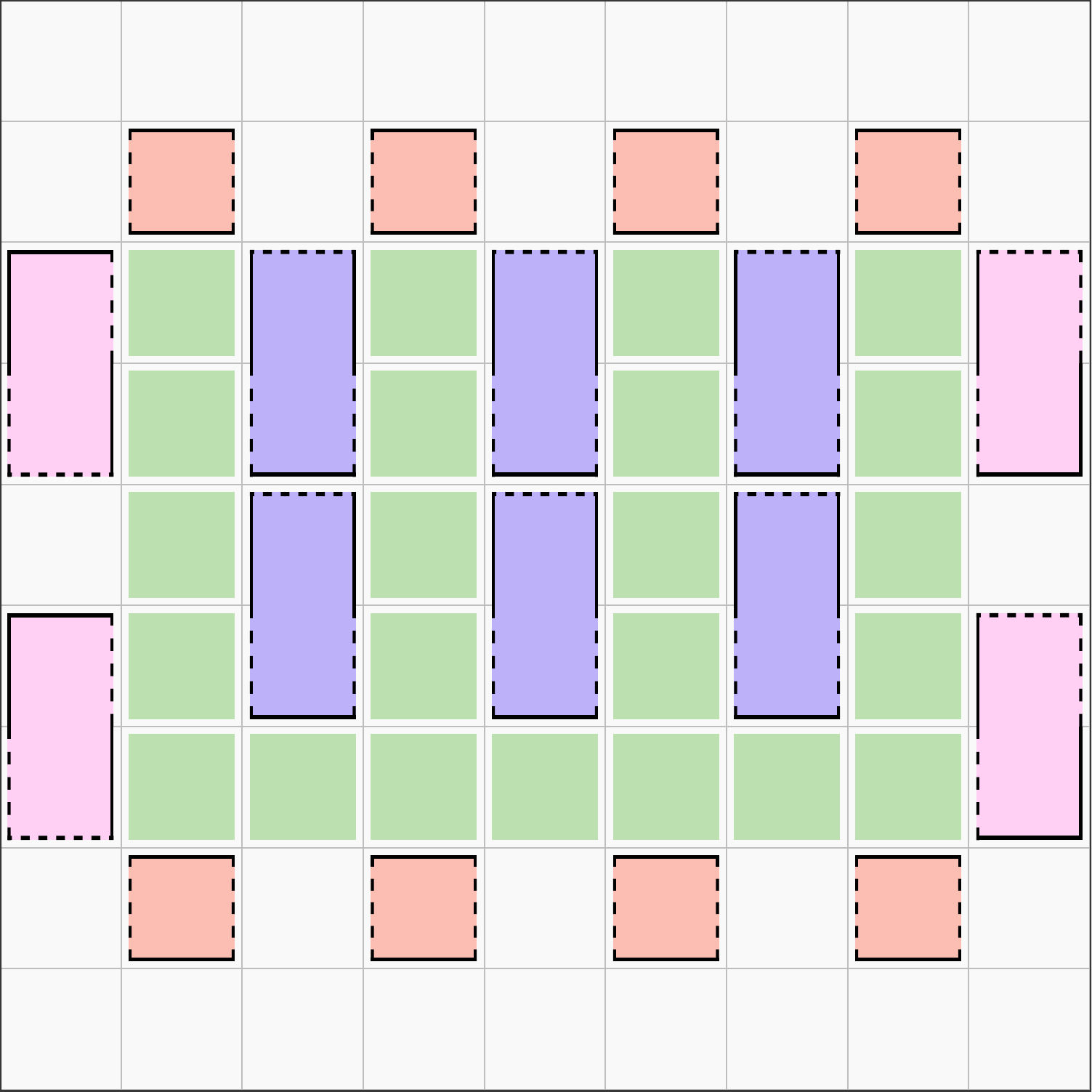} &  \hspace{\floatsep}
    & \includegraphics[scale=0.15]{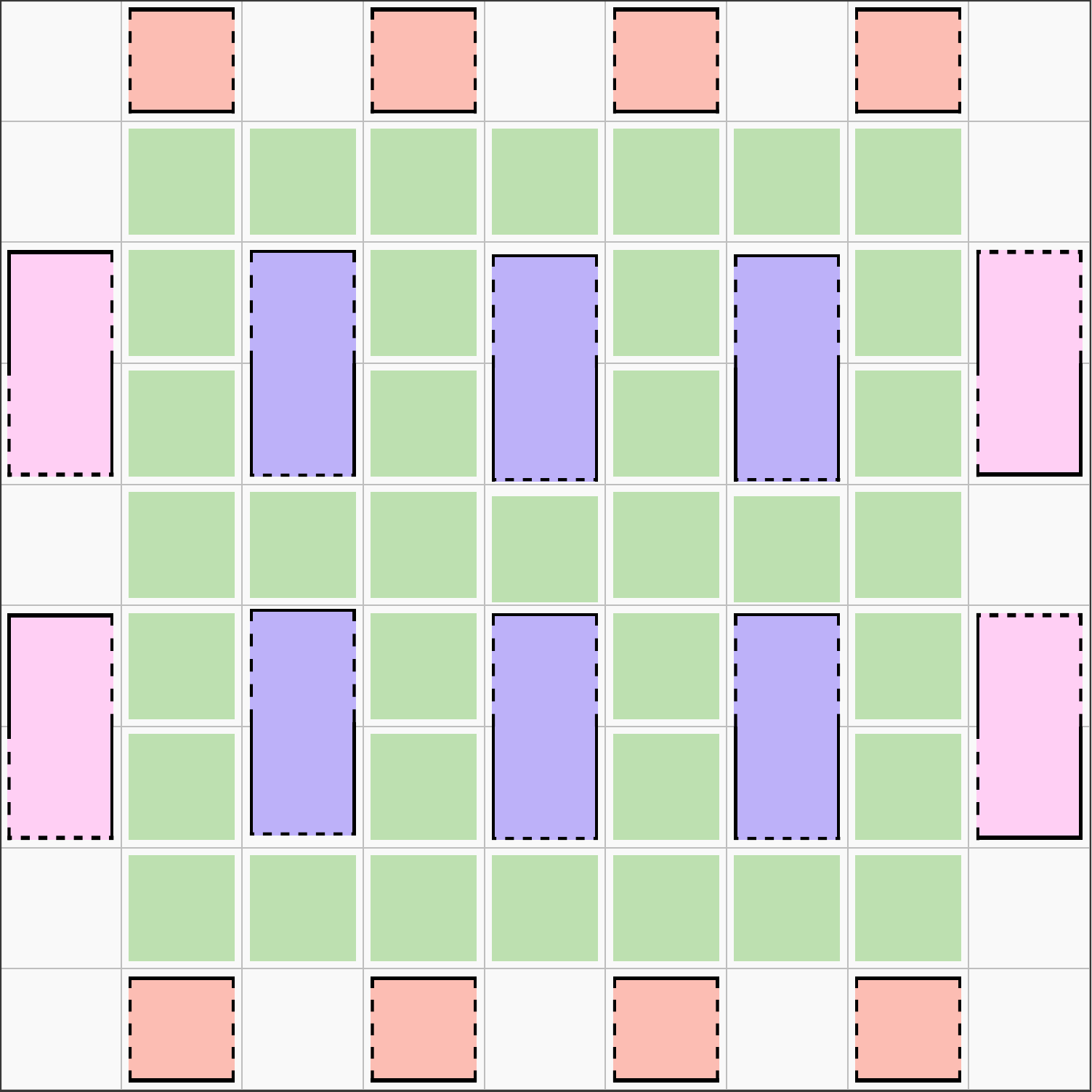} \\[\abovecaptionskip]
    \small (a) Compact & & \small (b) Parallelizable
  \end{tabular}
\caption{Examples of central zones comprising data qubits surrounded by bus qubits, and with ancillary and magic state storage qubits located at the boundary of the central zone. Given a layout of type (a) with $A$ aisles of data qubits and $P$ data qubit patches in each aisle, a modification can be done to transform it into a layout of type (b) by adding $P(2A+1)$ extra bus qubit tiles. Layout (b) facilitates the parallelization of multi-qubit measurements, as qubits can be connected using multiple paths.}
\label{fig:surface_code_layout}
\end{figure}

\cref{fig:surface_code_layout} illustrates central zones comprising data qubits surrounded by a quantum bus. Magic state storage qubits may be located anywhere at the boundary of a central zone, assuming that magic state distillation is performed externally. Similarly, ancillary qubits are located around the central zone, although this is not a constraint in our models. In \cref{fig:surface_code_layout}(a), we show the fast block layout proposed in \cite{litinski2018game}, while in \cref{fig:surface_code_layout}(b), we propose a slight modification to Litinski's layout by creating aisles of bus qubits between the data qubit patches and adding a top aisle to the layout to facilitate the parallelization of multi-qubit measurements, as qubits can be connected using multiple paths. Note that, for similar layouts in which multiple qubits are encoded in a number of tiles, Litiski's layout in which two qubits are encoded in two tiles is essentially optimal due to the fact that the number of qubits is related to the size of the boundary.

\section{The lattice surgery scheduling problem} \label{sec:problem}

The LSSP is defined with two necessary inputs: a quantum circuit with Pauli operations and a logical map of the layout. The input quantum circuit is given by a sequence of $m$ Pauli operations $\mathcal{R} = \{R_1, R_2, \ldots, R_{m}\}$ on $N$ qubits considering their order. Each operation $R \in \mathcal{R}$ is characterized by an angle (or measurement) from the set $\{\pm\pi/4, \pm\pi/8, \mathrm{M}\}$, representing $\pi/4$ rotations, $\pi/8$ rotations, and measurement operations, respectively, as well as a Pauli string of length $N$ (e.g., an assignment of a single-qubit Pauli element to each qubit $n \in \{1, 2, \ldots, N\}$). Let $\mathcal{R}^{\pi/4}$ denote the set of $\pm \pi/4$ rotations in $\mathcal{R}$ and $\mathcal{R}^{\pi/8}$ be the $\pm\pi/8$ counterpart. We use $R_{in}$ to represent the single-qubit Pauli operator used by operation $R_i$ on qubit $n$. Therefore, a rotation $R_i$ requires a qubit $n$ if $R_{in} \neq I$.

In quantum circuits, dependency constraints dictate the order in which quantum operations must be performed. Operations are independent if there is no precedence relationship between them according to the logical constraints of the circuit. Let a dependency check function $c: \mathcal{R} \times \mathcal{R} \rightarrow \mathbb{B}$ be used to verify whether a pair of rotations is independent. If $c(R_i, R_j) = 1$ for operations \mbox{$R_i, R_j \in \mathcal{R}, i \le j$,} then $R_i$ must be completed before starting $R_j$, that is, $R_j$ depends on $R_i$. The rules defining the dependency check function are discussed in \cref{sec:dependency_constraints}. Here, it suffices to state that the dependency constraints can be abstracted into a \textit{dependency graph} $\mathcal{G}_{\mathrm{dep}} = (\mathcal{M}, \mathcal{A})$, where the nodes $\mathcal{M}$ represent operations. A directed arc $a_{ij} \in \mathcal{A}$ if $c(R_i, R_j) = 1$.

\begin{figure}[t]
  \centering
  \begin{tabular}{ccc}
    \includegraphics[width=0.37\linewidth]{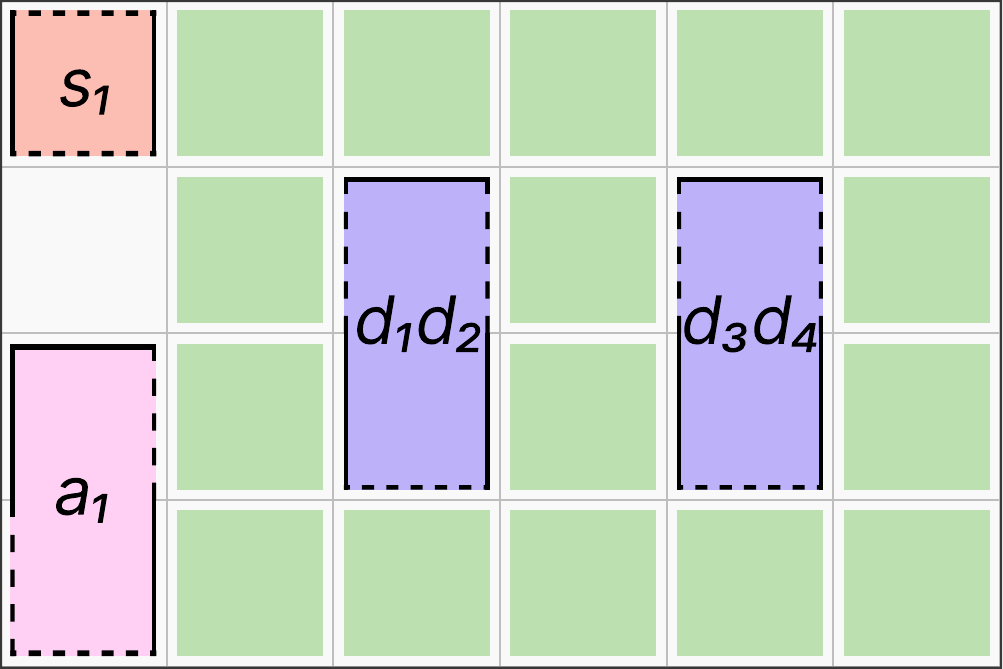} &
    \hspace{1cm} &
    \includegraphics[width=0.4\linewidth]{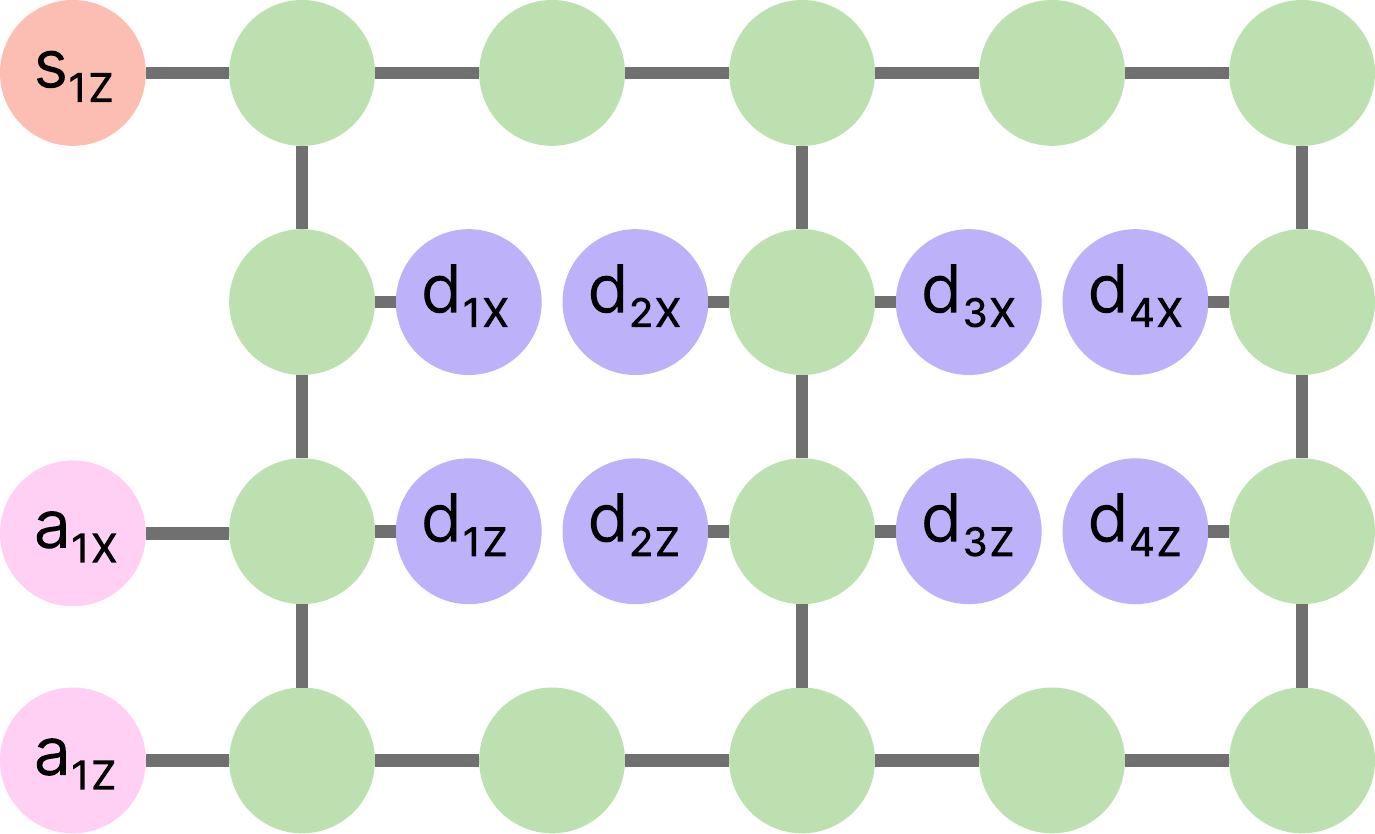}
    \\[\abovecaptionskip]
    \small (a) Logical qubit layout & & \small (b) Adjacency graph
  \end{tabular}
  \caption{Example layout for (a) the logical resources in the central zone converted to (b) an adjacency graph. Qubit patches and their accessible Pauli operators are converted into vertices in the adjacency graph, and edges represent the adjacencies between patches. The operators $Z_iZ_j$ (top) and $X_iX_j$ (bottom) of the two-tile, two-qubit patches can be represented by their own vertices, but additional constraints would be required to be added to our models to account for the choice of vertices to use when there is a possibility of connecting the ancilla patch to these operators. We therefore disregard these operators to simplify the generation of the ancilla patches.}
  \label{fig:layout_to_graph}
\end{figure}

The LSSP takes a logical map of the surface code layout as another necessary input, which specifies the resources available for running the circuit at the logical level. The layout can be abstracted into an undirected graph $\mathcal{G}_{\mathrm{adj}} = (\mathcal{V}, \mathcal{E})$, called the \textit{adjacency graph}, representing the logical resources, where the vertices $\mathcal{V}$ represent logical qubit patches and the edges $\mathcal{E}$ connect adjacent vertices in the lattice. Vertices are classified according to the type of qubit patch associated with them according to $\mathcal{V} = \{\mathcal{V}^\mathrm{B}, \mathcal{V}^\mathrm{D}, \mathcal{V}^\mathrm{S}, \mathcal{V}^\mathrm{A}\}$, where each vertex type is associated with the qubit types bus (B), data (D), magic state storage (S), or ancillary (A). \cref{fig:layout_to_graph} shows an example of the qubit adjacencies represented by an adjacency graph. It should be noted that the assignment of the qubits required by the circuit to the logical data qubits in the hardware should be known before solving the LSSP. Related studies usually randomly assign qubits to patches \cite{beverland2022surface}. However, more-elaborate methods for qubit placement may be used, such as minimizing qubit communication overhead by solving a variant of the quadratic assignment problem~\cite{lao2018mapping}.

A solution for the scheduling problem is represented by a sequence of time-ordered sets of operations \mbox{$\mathcal{T} = \{t_1, t_2, \ldots, t_{T}\}$,} where $T$ represents the number of time steps for executing all scheduled operations and each $t \in \mathcal{T}$ is the set of operations scheduled at the respective time step. The duration of a time step is defined by the longest operation scheduled for that time step, measured in logical cycles. As all operations represented by Pauli rotations take one logical cycle, the expected duration of each time step $i$ is equal to one logical cycle regardless of the number of operations scheduled in parallel within $i$, and the expected number of logical cycles $\mathbb{E}(N)$ to run all time steps is equal to $T$. The expected number of logical cycles $\mathbb{E}(N)$ can be converted into a concrete time measure, for example, in seconds, by using the wall-clock time of logical cycles.

The LSSP can now be formally stated as follows. Given a dependency graph $\mathcal{G}_{\mathrm{dep}}$ and an adjacency graph $\mathcal{G}_{\mathrm{adj}}$, for any rotation $R \in \mathcal{R}$, the LSSP seeks the time step at which $R$ should be performed to minimize the expected number of total logical cycles $\mathbb{E}(N)$.
The LSSP involves making two decisions: sequencing the operations and defining the resources at the logical level needed to perform each operation. We decompose the LSSP based on these decisions, where the sequencing decisions comprise the main problem while the resource usage decisions comprise subproblems to be solved at each time step.

\subsection{The primary problem} \label{sec:main_problem}

Let a \textit{pack} be a set of mutually independent logical operations that meet the layout constraints, such as those defined in \cref{sec:surface_code}. In other words, a pack $p = \mathcal{\widehat{R}} \subseteq \mathcal{R}$, where $c(R_i,R_j) = 0, \,\, \forall R_i,R_j \in \mathcal{\widehat{R}}$, and $p$ is a valid solution for the subproblem defined in \cref{sec:sub_problem} given the adjacency graph $\mathcal{G}_{\mathrm{adj}}$. Let $\mathcal{P}$ be the collection of all packs. For any pack $p \in \mathcal{P}$, we define a coefficient $A_{ip} = 1$ if rotation $R_i \in p$, or 0 otherwise. The mathematical formulation that we introduce concerns selecting the optimal combination of packs $\mathcal{P}^* \subseteq \mathcal{P}$ such that each operation is covered exactly once. Thus, for any pack $p \in \mathcal{P}$, the decision variable $x_p = 1$ if $p \in \mathcal{P}^*$, or 0 otherwise.

In the LSSP, a pack $p_i$ must be scheduled before another pack $p_j$ if there exists a pair of rotations $R_i \in p_i, R_j \in p_j$ such that $c(R_i,R_j)=1$. However, if there is also any $c(R_j,R_i)=1$, then a contradiction exists as this implies that $p_j$ must also be scheduled before $p_i$. For a set of packs $\mathcal{P}' \subseteq \mathcal{P}$, consider a graph where nodes are the packs in $\mathcal{P}'$ and arcs are the precedence relationships defined. A contradiction exists if a tour in the graph can be found between any subset of packs in $\mathcal{P}'$. Therefore, a feasible solution for the LSSP must not involve such a contradiction. Let us define $\widehat{\mathcal{P}}$ as all collections of packs in $\mathcal{P}$ that have a contradiction. A mathematical formulation for the LSSP is defined as follows:
\begin{align}
    \min \quad &\sum_{p\in \mathcal{P}}x_p\label{eq:obj_function_in_formulation}\\
    \text{s.t.}\quad & \sum_{p\in \mathcal{P}}A_{ip}x_p = 1, \quad \forall R_i \in \mathcal{R}, \label{cstr:cover_rotation}\\
    &\sum_{p\in \eta} x_p \leq |\eta| - 1, \quad \forall \eta \in \widehat{\mathcal{P}}, \label{cstr:subtour_cut}\\
    &x_p \in \{0,1\}, \quad \forall p \in \mathcal{P}. \label{cstr:domain}
\end{align}
The objective function~\eqref{eq:obj_function_in_formulation} is defined such that it minimizes the number of packs chosen given that each pack requires the same amount of time to be executed. Constraints~\eqref{cstr:cover_rotation} impose that each rotation must be scheduled exactly once. Constraints~\eqref{cstr:subtour_cut}, where $\eta$ refers to packs with a contradiction, ensure that the packs selected will not involve a contradiction by eliminating the formation of tours in the selected packs. Finally, constraints~\eqref{cstr:domain} define the domain of the decision variables.

\subsection{The subproblem} \label{sec:sub_problem}

Defining the set of valid packs $\mathcal{P}$ for the main problem requires solving a packing subproblem. This subproblem involves checking if a set of mutually independent operations can be performed in parallel by verifying whether ancilla patches can be generated on the topological code layout to connect the qubits required by the operations without violating the layout constraints. While checking if a given packing is feasible is enough to define a valid pack for the main problem, we model the packing subproblem such that the usage of logical resources is minimized to reduce the error rate for the operation.

Given $\mathcal{\widehat{R}}$, a set of mutually commuting rotations, where \mbox{$\mathcal{\widehat{R}}^{\pi/8}, \mathcal{\widehat{R}}^{\pi/4} \subseteq \mathcal{R}'$} represent the subsets of $\pi/8$ and $\pi/4$ rotations, respectively, let $\mathcal{V}^\mathrm{D}_i$ be the set of data qubit vertices required by the rotation $R_i$ in the adjacency graph $\mathcal{G}_{\mathrm{adj}} = (\mathcal{V}, \mathcal{E})$. Let the set of incident edges to a vertex $v \in \mathcal{V}$ be defined as \mbox{$\delta(v) = \{(i,j) \mid i = v \vee j = v, \forall (i,j) \in \mathcal{E}\}$.} For any edge $e \in \mathcal{E}$ and rotation $R_i \in \mathcal{\widehat{R}}$, we define a set of decision variables specifying the assignment of the edges as $y_e^{i} = 1$ if the ancilla patch for $R_i$ uses edge $e$, or 0 otherwise. Similarly, for any bus qubit $v\in \mathcal{V}^\mathrm{B}$, another set of decision variables is defined as $z^{i}_v = 1$ if $R_i$ uses vertex $v$, or 0 otherwise. The packing subproblem is then defined as follows:

\begin{align}
    \min \quad & \sum_{v\in \mathcal{V}^\mathrm{B}}\sum_{R_i\in \mathcal{\widehat{R}}}z^i_v \label{obj:sub_problem}\\
    \text{s.t.} \quad &\sum_{e\in \delta(v)} y^i_e =1, \quad \forall v\in \mathcal{V}^\mathrm{D}_{i}, R_i \in \mathcal{\widehat{R}}, \label{cstr:sub_problem_data_qubit_assignment}\\
    &\sum_{e\in \delta(v)} y^i_e =1, \quad \forall v\in \mathcal{V}^\mathrm{S}, R_i \in \mathcal{\widehat{R}}^{\pi/8}, \label{cstr:sub_problem_ms_assignment}\\
    &\sum_{e\in \delta(v)} y^i_e =1, \quad \forall v\in \mathcal{V}^\mathrm{A}, R_i \in \mathcal{\widehat{R}}^{\pi/4},\label{cstr:sub_problem_zs_assignment}\\
    &\sum_{R_i\in \mathcal{\widehat{R}}}z_v^i \leq 1, \quad \forall v\in \mathcal{V}^\mathrm{B}, \label{cstr:sub_problem_avail_bus_qubit}\\
    &2y^i_{(j,k)}\leq z^i_j + z^i_k, \quad \forall (j,k)\in \mathcal{E}, R_i \in \mathcal{\widehat{R}}, \label{cstr:sub_problem_z_y_connectivity}\\
    &\sum_{e\in \mathcal{E}}y_e^i = \sum_{v\in \mathcal{V}}z^i_v -1, \quad \forall R_i\in \mathcal{\widehat{R}}, \label{cstr:sub_problem_acyclicity}\\
    &y^i_e, z^i_v \in \{0,1\}, \quad \forall e\in \mathcal{E}, v\in \mathcal{V}, R_i\in \mathcal{\widehat{R}} ,\label{cstr:sub_problem_domain}
\end{align}
The objective function~\eqref{obj:sub_problem} minimizes the total number of bus qubits used to build the ancilla patches. Constraints~\eqref{cstr:sub_problem_data_qubit_assignment} ensure that all data qubits required by each rotation are connected to the ancilla patch generated for that rotation by a single edge, which guarantees that no data qubit required is crossed by the ancilla patch. Constraints~\eqref{cstr:sub_problem_ms_assignment} and \eqref{cstr:sub_problem_zs_assignment} indicate that there must be exactly one magic state storage qubit and ancillary qubit connected to the $\pi/8$ and $\pi/4$ rotations, respectively. Constraints~\eqref{cstr:sub_problem_avail_bus_qubit} impose that a bus qubit can be used by no more than one ancilla patch. Constraints \eqref{cstr:sub_problem_z_y_connectivity} ensure the connection between the two variable sets by stating that if the edge $(j,k)$ is used by any ancilla patch, then vertices $j$ and $k$ are also used by it. Constraints~\eqref{cstr:sub_problem_acyclicity} guarantee that the ancilla patch generated is a tree. One property of trees is that the number of edges they contain is equal to their number of vertices minus one. The vertices of ancilla patches are composed of the required qubits and all bus qubits used to connect them. Finally, constraints~\eqref{cstr:sub_problem_domain} define the domain of the binary decision variables.

Enumerating all $\mathcal{O}(2^m)$ possible packs and solving the packing subproblem for all of them is usually impractical. Although it is possible to address the main problem and the enumeration of the contradicting packs for constraints~\eqref{cstr:subtour_cut} in real time by using techniques like column generation or cutting-plane algorithms, considering the scale of real quantum circuits, solving the main problem exactly is also impractical. Therefore, we next present an algorithm that schedules operations using a scalable heuristic.

\section{A greedy approach to lattice surgery scheduling} \label{sec:method}

To address the LSSP, a crucial step involves building the dependency graph for searching for valid packs. This requires a fast algorithm to perform dependency checks while preserving the true dependency of operations.

Scheduling a long-range multi-qubit operation requires the generation of an ancilla patch connecting the multiple qubits required, called \textit{terminals}. Although any ancilla patch meeting the connectivity requirement is valid, smaller patches are desired as they result in lower error rates. While connecting pairs of qubits is computationally easy, as it requires solving a shortest path problem, connecting to more terminals introduces the NP-hard terminal Steiner tree problem~\cite{lin2002terminal, drake2004approximation}, as the connected qubits are required to be leaf nodes in a tree. Due to the hardness of generating Steiner trees, heuristics are often employed to quickly find near-optimal solutions~\cite{ribeiro2002hybrid, uchoa2012fast, fischetti2017thinning, pajor2018robust, ljubic2021solving}. Scaling tree generation for potentially millions of operations is essential.

Operations can be scheduled in parallel with the aim of reducing the expected circuit execution time. Checking the feasibility of scheduling multiple operations in parallel poses a challenge,  since it requires efficiently packing trees into the adjacency graph. This problem is related to the Steiner forest packing problem, which, along with the Steiner tree packing problem, has been extensively investigated~\cite{lau2005packing, gassner2010steiner, hoang2012steiner, braunstein2018cavity, sun2022packing}.

To tackle these challenges, we have developed fast heuristics for creating the dependency graph, searching for Steiner trees, and packing multiple trees in parallel. The LSSP is solved using an earliest-available-first (EAF) algorithm based on the EAF policy, where operations are scheduled as they become available, given the dependency constraints. We employ a greedy heuristic to solve the forest packing problem, maximizing operations packed among the available candidates. This process is repeated until all operations have been scheduled, considering layout constraints and updated dependency graphs. \cref{alg:scheduler} outlines the EAF algorithm for the LSSP, providing a high-level view of our designed approach. Further details on its steps are presented below.

\begin{algorithm}[H]
\footnotesize
\caption{~Earliest-Available-First Scheduling Algorithm}
\label{alg:scheduler}
\begin{algorithmic}[1]
\STATE \textbf{input} Pauli rotation circuit and adjacency graph;
\STATE Identify dependencies among operations to build a dependency graph;
\WHILE {dependency graph is not empty}
\STATE \textit{Candidates} $\leftarrow$ root nodes of the dependency graph;
\STATE Solve the forest packing problem for the candidates, considering the adjacency graph;
\STATE Remove scheduled operations from the dependency graph;
\ENDWHILE
\RETURN operations schedule and trees generated for multi-qubit operations
\end{algorithmic}
\end{algorithm}

\subsection{Dependency graph generation} \label{sec:dependency_constraints}

The order in which operations appear in a quantum circuit determines their dependency relationships; in general, operations must be scheduled following this order. However, the qubits and the Pauli operators required by the operations may allow some operations to commute with others, meaning that they can be applied in any order without affecting the final quantum state. This creates an opportunity to schedule commuting operations in parallel considering their relative positions within the circuit. This section explores different methods for generating dependency constraints, each of which has its own advantages and trade-offs.

As previously stated, the dependency constraints can be abstracted into a dependency graph $\mathcal{G}_{\mathrm{dep}} = (\mathcal{V},\mathcal{A})$. The vertices, which represent quantum operations, are divided into the subsets $\{\mathcal{V}^{\pi/8}$, $\mathcal{V}^{\pi/4}$, and $\mathcal{V}^{\mathrm{M}}\}$, where each vertex type is associated with a quantum operation type among the $\pi/8$ rotation, the $\pi/4$ rotation, and the qubit measurement (M).

The commutation check for Pauli rotations is defined as follows based on the symplectic representation of Pauli operators as shown in
\cref{app:improved_gate_optimization}.  Given two Pauli operators with symplectic representations $P =(\theta_P|x_P|z_P)$ and $Q=(\theta_Q|x_Q|z_Q)$, $P$ and $Q$ commute if $x_P\cdot z_Q + x_Q\cdot z_P\mod 2 = 0$.

A straightforward way to generate the dependency graph following the definition above, which we refer to as the \textit{general rule}, is by visiting each pair of nodes and verifying whether they commute, which requires $\mathcal{O}(|\mathcal{V}|^2$) commutation checks. Whenever a pair of nodes $i$ and $j$ does not commute, then an arc $a_{ij}$ is added to the graph. The dependency graph generated through commutation checks is a directed acyclic graph. Therefore, it can be generated in a transitive reduced form by avoiding redundant arcs that are inferred from the existing arcs in the graph while preserving the relationships of connectivity between nodes. Thus, if $a_{ij}, a_{jk} \in \mathcal{A}$, then a dependency constraint exists between $i$ and $k$ regardless of their commutativity, that is, the dependency check function $c(i,k) = 1$. This graph can be generated in a transitive reduced form using a depth-first search algorithm for the commutation checks. In this way, the number of commutation checks is reduced by $\mathcal{O}(|\mathcal{A}|$).

\begin{figure}
  \centering
  \begin{tabular}{cccccc}
    \begin{tikzpicture}[node distance={15mm}, main/.style={circle, draw}]
        \node[main] (1) {$R_1$};
        \node[main] (2) [below left of=1] {$R_2$};
        \node[main] (3) [below right of=1] {$R_3$};
        \node[main] (4) [below right of=2] {$R_4$};

        \draw [->] (1) -- (2);
        \draw [->] (1) -- (3);
        \draw [->] (3) -- (4);
    \end{tikzpicture} & &
    \begin{tikzpicture}[node distance={15mm}, main/.style={circle, draw}]
        \node[main] (1) {$R_1$};
        \node[main] (2) [below left of=1] {$R_2$};
        \node[main] (3) [below right of=1] {$R_3$};
        \node[main] (4) [below right of=2] {$R_4$};

        \draw [->] (1) -- (2);
        \draw [->] (2) -- (3);
        \draw [->] (3) -- (4);
    \end{tikzpicture} & &
    \begin{tikzpicture}[node distance={15mm}, main/.style={circle, draw}]
        \node[main] (1) {$R_1$};
        \node[main] (2) [below left of=1] {$R_2$};
        \node[main] (3) [below right of=1] {$R_3$};
        \node[main] (4) [below right of=2] {$R_4$};

        \draw [->] (1) -- (2);
        \draw [->] (1) -- (3);
        \draw [->] (2) -- (4);
        \draw [->] (3) -- (4);
    \end{tikzpicture}\\[\abovecaptionskip]
    %\small (a) Full & &
    \small (a) General & & \small (b) Serial & & \small (c) Trivial
  \end{tabular}

  \caption{Dependency graphs generated for a circuit with four operations and four qubits, where \mbox{$R_1 = \{I,X,Y,I\}$,} $R_2 = \{Z,I,Z,I\}$, $R_3 = \{I,Y,I,Y\}$, and $R_4 = \{X,X,X,Y\}$. Even though $R_1$ and $R_4$ do not commute, the graph for the general rule (a) has only non-redundant arcs. The graph for the serial rule (b) has a single path that respects the order of the operations in the circuit. The graph for the trivial rule (c) restricts commutation and provides a faster way to approximate the general graph.}
  \label{fig:dependency_graphs}
\end{figure}
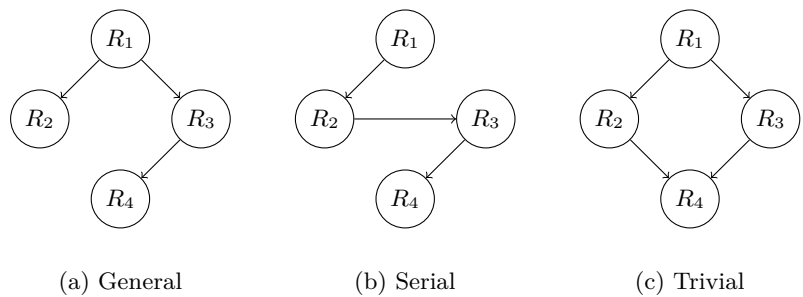

Other rules can be employed to generate dependency graphs that impose dependency constraints, even if operations commute. The \textit{serial rule} guarantees that operations are scheduled sequentially and only requires $\mathcal{O}(|\mathcal{V}|$) operations to be generated by adding the arcs $a_{i(i+1)}$, \mbox{$\forall i \in \{1, \ldots, |\mathcal{V}|-1\}$.} Another rule, the \textit{trivial rule}, enforces commutation only when qubits required by one operation are disjoint from those required by the other. From the symplectic representation of rotations $R_i$ and $R_j$, trivial commutation is only possible when $(x_{R_i}\lor z_{R_i}) \cdot (x_{R_j}\lor z_{R_j}) = 0$. We note that if data qubits cannot contribute to different measurements simultaneously, as stated in a footnote in \cref{sec:surface_code}, the trivial rule results in the true dependency constraints for the scheduling, as operations commuting according to the general rule might violate this condition. \cref{fig:dependency_graphs} shows a comparison of the dependency graphs generated using the rules described for a small circuit. Different rules introduce some trade-offs in terms of parallelization potential, but the trivial rule proves to be scalable for circuits of a size suitable for practical applications, as discussed in \cref{sec:scalability_analysis}.

\subsection{Solving forest packing problems} \label{sec:forest_packing}

The LSSP is addressed by employing an EAF algorithm that prioritizes the scheduling of operations as early as possible. An operation becomes a candidate for being scheduled at the subsequent time step if it is not dependent on any other operation that has not yet been scheduled. This condition implies that only operations at the root node of the dependency graph are candidates for being scheduled to occur in the same time step. As the dependency graph is a directed acyclic graph, there must be always at least one node that meets the condition described unless all operations have been scheduled or when qubit availability is considered. In the latter case, waiting until all qubits required by at least one root node are available would circumvent this issue.

Given the set of candidate operations and the adjacency graph, the scheduling of the candidates requires solving a forest packing problem if at least one multi-qubit operation is a candidate, where each tree to be packed represents the ancilla patch to be used to generate the multi-qubit entanglements. \cref{alg:tree_packing} presents the greedy algorithm designed to solve the forest packing problem. For the sake of brevity, we describe the greedy algorithm implemented only for the scheduling of $\pi/8$ rotations, but the steps are applicable for $\pi/4$ rotations.

\begin{algorithm}[H]
\footnotesize
\caption{~Greedy Forest Packing Algorithm}
\label{alg:tree_packing}
\begin{algorithmic}[1]
\STATE \textbf{input} set of candidate operations ($\mathcal{\widehat{R}}$) and adjacency graph ($\mathcal{G}_{\mathrm{adj}} = (\mathcal{V}, \mathcal{E})$);
\STATE Initialize forest packed $\mathcal{S} \leftarrow \emptyset$;

\FORALL{$R_i \in \mathcal{\widehat{R}}$}
    \STATE Temporarily update $\mathcal{V} \leftarrow \mathcal{V}^{\mathrm{D}}_{i}$;
    \STATE Initialize current tree $\mathcal{G}_{i} = (\mathcal{V}^{\mathrm{D}}_{i}, \emptyset)$;

    \IF {$|\mathcal{V}^{\mathrm{D}}_{i}| > 1$}
        \STATE $\mathcal{G}_{i} \leftarrow $ the solution to the terminal Steiner tree problem connecting terminals $\mathcal{V}^{\mathrm{D}}_{i}$ in the graph $\mathcal{G}_{\mathrm{adj}}$;
    \ENDIF

    \IF {$\mathcal{G}_{i} \neq \emptyset$}
        \IF {$R_i \in \mathcal{V}^{\pi/8}$}
            \IF {$|\mathcal{V}^{\mathrm{D}}_{i}| > 1$}
                \STATE Replace $\mathcal{G}_{i}$ by a node $g$ in $\mathcal{G}_{\mathrm{adj}}$ \textbf{and} $s \leftarrow g$ \textbf{else} $s \leftarrow v, v \in \mathcal{V}^{\mathrm{D}}_{i}$;
            \ENDIF
            \STATE $m^* = \arg \min_{m \in \mathcal{V}^\mathrm{S}} d(s, m) $;
            \IF {$m^* = \emptyset$}
                \STATE Go to line 3;
            \ENDIF
            \STATE Update $\mathcal{G}_{i}$ with the path found to $m^*$;
            \STATE $\mathcal{V}^\mathrm{S} \leftarrow \mathcal{V}^\mathrm{S} \setminus \{m^*\}$;
        \ENDIF
        \STATE $\mathcal{S} \leftarrow \mathcal{S} \cup (R_i, \mathcal{G}_{i})$ \textbf{and} $\mathcal{V} \leftarrow \mathcal{V} \setminus \{\mathcal{V}^\mathrm{B}_{i}\}$;
    \ENDIF
\ENDFOR

\RETURN $\mathcal{S}$
\end{algorithmic}
\end{algorithm}

First, the set of candidate operations $\mathcal{\widehat{R}}$ and the adjacency graph $\mathcal{G}_{\mathrm{adj}} = (\mathcal{V}, \mathcal{E})$ are input (line~1). We initialize the forest packing set $\mathcal{S}$ as empty (line~2). The greedy algorithm tentatively schedules one random operation $R_i \in \mathcal{\widehat{R}}$ at a time (line~3). The data qubits $\mathcal{V}^{\mathrm{D}}_{i} \subset \mathcal{V}$ required by $R_i$ define the terminals to be used to generate each tree. For each candidate operation, we temporarily remove from $\mathcal{V}$ all data qubits not required by $R_i$ to avoid using them when generating the trees (line~4).

The forest packing problem requires that all trees generated are element-disjoint, meaning that they can share terminal vertices but not internal vertices or edges. Each tree must be a subgraph of $\mathcal{G}_{\mathrm{adj}}$. The tree generation may require two steps. First, if multiple terminals are required, a Steiner tree is generated to connect only the required terminals (lines~6--8). Next, if the candidate is a $\pi/8$ rotation, a vertex associated with a magic state storage qubit is added to the tree (lines~10--20).

The Steiner tree is generated in line~7 by solving the terminal Steiner tree problem to connect the terminals $\mathcal{V}^\mathrm{D}_{i}$ within the graph $\mathcal{G}_{\mathrm{adj}}$. We implement a modified version of Mehlhorn's algorithm~\cite{mehlhorn1988faster} for the terminal Steiner tree problem variant. Our algorithm generates a complete graph with all terminal vertices, where each edge represents the shortest path between the connected vertices, which is found using a bidirectional Dijsktra algorithm. Then, it finds the minimum spanning tree using the Kruskal algorithm to connect all vertices in the complete graph. The Steiner tree generated $\mathcal{G}_{i}$ is provided, connecting all shortest paths chosen for the minimum spanning tree. Since terminals are required to be leaves in the final tree, we ensure that only bus qubits are used as internal vertices in the shortest paths generated by temporarily disconnecting non-bus qubits from the quantum bus. If no feasible tree is found (line~9), then the algorithm moves on to the next candidate, as $R_i$ cannot be scheduled at this time step.

If the candidate $R_i$ is a $\pi/8$ rotation (line~10), one of the magic state storage qubits in $\mathcal{V}^\mathrm{S}$ must be added to the tree as a terminal. If this is the case, the magic state storage qubit chosen is the one that is closest to the tree previously generated or to the single qubit required, if this is the case. For the latter, we replace all vertices and edges used for the tree by a single vertex $g_{i}$ and associate this vertex with the source vertex $s$. Otherwise, $s$ is associated with the single qubit required by $R_i$ (lines~11--13). Then, we solve a shortest path problem for each magic state storage qubit $m \in \mathcal{V}^\mathrm{S}$ from the source $s$ to the magic state storage qubit $m$. Given the distance $d(s,m)$ between the source $s$ and a magic state storage qubit $m$, the magic state storage qubit chosen $m^*$ is the one closest to $s$ (line~14); however if no magic state storage qubit can be connected to the tree, this operation is skipped (lines~15--17). If a feasible connection is found, the algorithm connects the shortest path found to the tree (line~18) and removes $m^*$ from the set of storage qubits available (line~19).

Once a tree connecting all terminals is generated, this operation is considered to be scheduled at the current checked time step, by adding the operation $R_i$ and the tree $\mathcal{G}_i$. This process is repeated for another candidate operation using an updated version of the adjacency graph in which all bus qubits $\mathcal{G}^\mathrm{B}_{i}$ used by the set of trees generated in this time step are removed from the graph (line~21). Whenever a tree cannot be generated for a candidate, then the candidate is not scheduled at this time step and waits until the next one. Once all candidates have been checked, the algorithm returns the set of operations scheduled and the trees generated for each of them (line~24).

The scheduling generation is significantly sped up by caching solutions found during the process, such as those for the shortest path problem, for the terminal Steiner tree problem, and for the forest packing problem. Thus, whenever a problem arises for inputs that have already been checked, the previous solution can quickly be retrieved from the cache and checked if it is feasible for the current time step.

\section{Computational experiments} \label{sec:results}

This section describes the results of the extensive computational experiments we performed to test the greedy approach proposed to solve the LSSP. The experiments were run using the Google Cloud Platform with nodes comprising an Intel Xeon Gold 6268CL CPU with a 2.80~GHz clock and 256~GB of RAM, limited to a single computing core per run. We implemented the greedy scheduling algorithm in Python and used the NetworkX library for graph operations.

\subsection{Test circuits} \label{sec:test_circuits}

Two sets of circuits were used to test the implemented algorithms. The first set contains random circuits generated using a structure we defined, while the second contains circuits generated in the Clifford + $T$ basis to represent quantum circuits with real applications. In particular, we used Hamiltonian simulation of various systems as example circuits with real-world applications.

The random circuits were created using a scripted approach to follow the characteristics of circuits that emulate valid outputs of transpilation. An $N$-qubit circuit consists of many $\pi/8$ rotations requiring a specified number of qubits followed by $N$ final qubit measurements. Key characteristics considered during circuit generation include the circuit length $m$, the total number of qubits required $N$, and the average percentage of qubits involved in each operation~$N_{\%}$.

When sampling a rotation for a random circuit, the process begins by determining the number of qubits in its Pauli operator, drawn from a normal distribution $\mathscr{N}(N \times N_{\%}, 2)$. Subsequently, the qubits assigned to a Pauli operator according to the specified number of qubits required are randomly chosen. Then, a Pauli operators among $X$, $Y$, and $Z$ is assigned to the chosen qubits.
For our computational experiments, the random circuits were generated with combinations from the sets $m = \{ 10$$,$$000$, $20$$,$$000$, $30$$,$$000\}$, $N = \{10,30,50\}$, and $N_{\%} = \{0.15,0.50,0.85\}$. This totals 27 combinations, where each was generated by five seeds, resulting in 135 random circuits.

The application-inspired circuits followed well-known quantum algorithms for simulating quantum dynamics. They were generated using Hamiltonian simulation for a single step of a Suzuki--Trotter decomposition. Five types of Hamiltonians were generated: electronic structure Hamiltonians of interest in quantum chemistry, one-dimensional chains of random Pauli interactions, transverse-field Sherrington--Kirkpatrick model (TFSK) Hamiltonians, rotated surface code (RSC) Hamiltonians, and one-dimensional Heisenberg XYZ Hamiltonians. For quantum chemistry circuits, we used the electronic structure Hamiltonians of the molecules H$_2$, LiH, and H$_2$O. The representations of the Hamiltonians for these molecules were generated using Tangelo~\cite{senicourt2022tangelo}, an open source Python module for quantum chemistry. We implemented a single step of a Trotter decomposition using PennyLane~\cite{arrazola2023differentiable}, another open source Python module for quantum algorithms, ensuring an error rate below $10^{-5}$. The circuits for one-dimensional chains of random Pauli interactions were generated by randomly selecting Pauli interactions between all sets of three neighbouring qubits, with each interaction having a different random interaction strength between $-$1 and 1, along with a fixed $X$-type interaction on each qubit of strength 1, with an evolution time of 1. We considered one-dimensional chains with 10, 20, 30, 40, and 50 qubits. The TFSK Hamiltonian circuits involved $X$-type interactions of strength 1 on all qubits and $ZZ$ interactions between all pairs of qubits with interaction strength randomly chosen between $-$1 and 1, and evolved for a time of 1. We explored 10- and 15-qubit circuits. The circuits using RSC Hamiltonians comprise all the check operators of the RSC of distances 3, 5, and 7, with each of the $X$ and $Z$ check operators given a uniformly random weight between 0 and 1, and evolved for a time of 1. The one-dimensional Heisenberg XYZ Hamiltonians consisted of Heisenberg XYZ interactions, with weights for the $XX$, $YY$, and $ZZ$ terms randomly chosen using an uniform distribution between $-$1 and 1, and were evolved for a time of 1. One-dimensional chains with 5, 10, 20, 30, and 40 qubits were considered. All these circuits were decomposed using a Solovay--Kitaev algorithm that decomposes gates with arbitrary rotations into Clifford + $T$ gates, with at $L^2$-norm error of at most $2.5\times 10^{-2}$. They were then converted to Pauli rotations using the rules shown in \cref{fig:basis_conversion} and optimized using the transpilation algorithm described in \cref{app:improved_gate_optimization}. A table summarizing the characteristics of these circuits, before and after the transpilation, is given in \cref{app:real_circuits_summary}.

\subsection{Analysis of dependency graph generation rules} \label{sec:scalability_analysis}

In this round of experiments, we solved the LSSP using the dependency graph generation rules described in \cref{sec:dependency_constraints}. The layouts considered were generated following the architecture presented in \cref{sec:surface_code}, with the number of data qubits equal to $N$ and a fixed number of magic state storage qubits $|\mathcal{V}^\mathrm{S}| = 3$ located around the central zone. The percentage gap used to compare solutions is defined as $gap = 100(S - S^*)/S^*$, for a solution $S$ compared to another solution $S^*$.

Before presenting the results of our experiments, we define a metric to analyze the dependency graphs generated by each rule. Let us denote the depth of a node $i$ in the dependency graph \mbox{$\mathcal{G}_{\mathrm{dep}} = (\mathcal{N},\mathcal{A})$} as $D(i) = \max_{(i,j) \in \mathcal{A}} [D(j) + 1]$, where $(i,j) \in \mathcal{A}$ indicates a directed arc from node $i$ to node $j$. Then, the width $W_d$ of $\mathcal{G}_{\mathrm{dep}}$ at the depth $d$ is defined as the total number of nodes with depth $d$ in $\mathcal{G}_{\mathrm{dep}}$, that is, $W_d = |\{i \in \mathcal{N} | D(i) = d\}|$. Given that $D_{\mathrm{max}} = \max_{i \in \mathcal{N}} D(i)$ is the maximum depth of $\mathcal{G}_{\mathrm{dep}}$, we define the average width of $\mathcal{G}_{\mathrm{dep}}$ as
\begin{equation}
    \overline{W} = \dfrac{1}{D_{\mathrm{max}}} \sum\limits_{d = 1}^{D_{\mathrm{max}}} W_d. \label{eq:avg_width}
\end{equation}
\noindent In other words, if we were to relax the layout constraints, $D(i)$ would denote at which time step the operation $i$ is scheduled, $W_d$ would be the number of operations scheduled for time step $d$, and $\overline{W}$ would represent the average number of operations scheduled per time step. Based on these definitions, $\mathrm{max}(D) = \max_{i \in \mathcal{R}} D(i)$ represents the circuit depth and is a lower bound for the number of time steps in the optimal solution of the LSSP. In addition, $\overline{W}$ is a measure of the parallelization potential of a circuit. Therefore, for the serial rule (see \cref{sec:dependency_constraints}), $\overline{W} = 1$ and $\mathrm{max}(D) = \mathcal{R}$, that is, the circuit depth is equal to the circuit length.

\begin{figure}[h]
  \centering
    \includegraphics[width=0.5\linewidth]{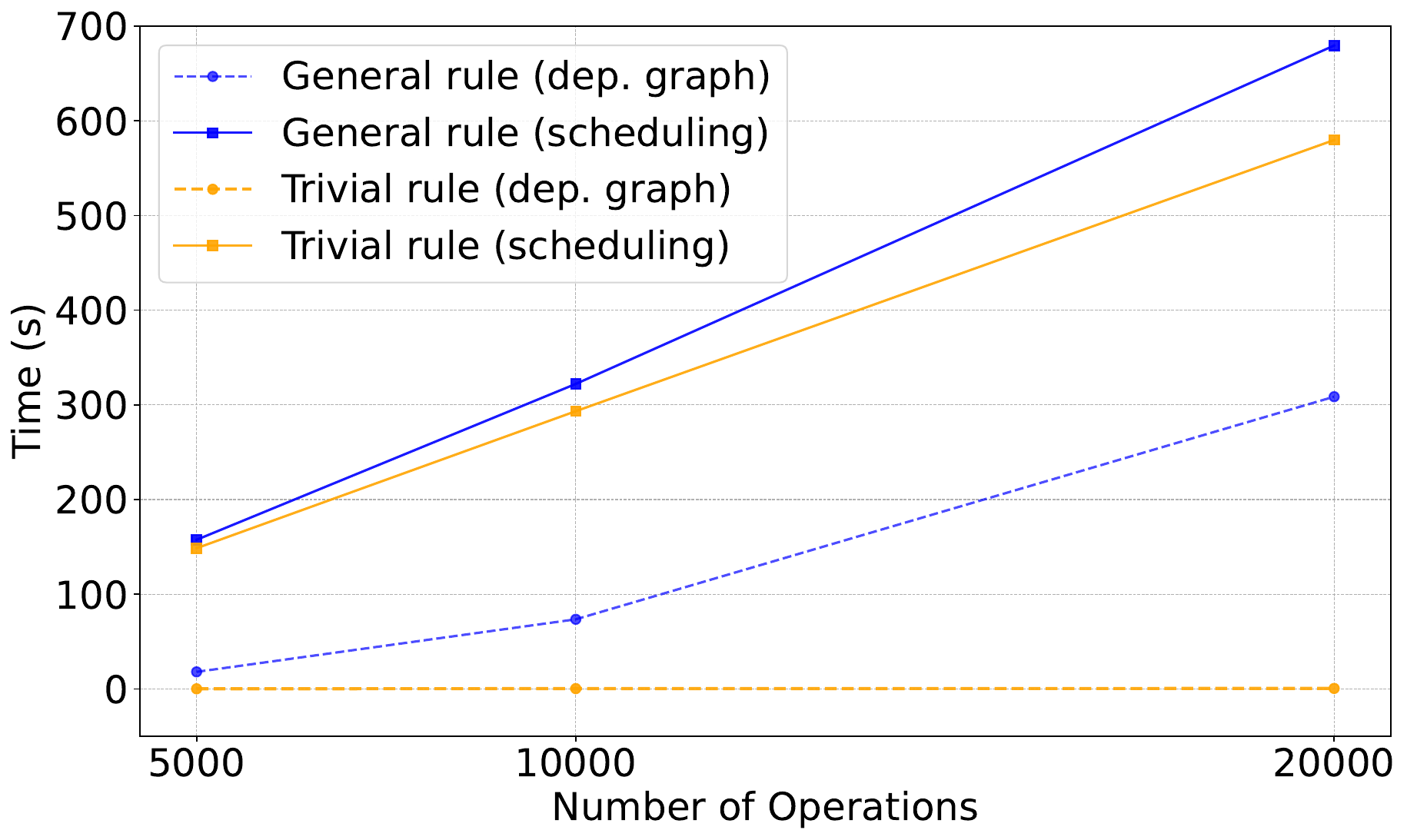}
  \caption{Time needed, in seconds, to generate the dependency graph and the schedule versus circuit size, for the general rule and the trivial rule. The time to generate the dependency graph using the trivial rule does not scale with the number of operations, making it particularly suitable for large-scale circuits.}
  \label{fig:commutation_scalability}
\end{figure}

\cref{tab:avg_values_commutation_rule_analysis} displays the average results obtained in the set of experiments for each rule used to generate the dependency graph. We note that serial scheduling can be considered an upper bound for the LSSP. Complementary data generated from this round of experiments are presented in \cref{tab:gap_to_serial_scheduling} and \cref{fig:commutation_scalability}.

\begin{table}
    \caption{\label{tab:avg_values_commutation_rule_analysis} Average statistical values for the experiments performed using each dependency graph generation rule. $\overline{W}$: the average dependency graph width; $t_{\mathrm{dep}}$: the dependency graph generation time, in seconds;  $t_{\mathrm{sch}}$: the scheduling time using the earliest-available-first algorithm, in seconds; $t_{\mathrm{tot}}$: the total time ($t_{\mathrm{dep}} + t_{\mathrm{sch}}$); $gap$: the percentage gap of an LSSP solution $S$ for a circuit compared to the best solution among the three rules $S^*$. Times are not reported for the serial rule, as solutions for the LSSP found after the dependency graph is generated using the serial rule are trivially determined by multiplying the circuit depth by the expected time needed to execute each operation.}
    \centering
    \begin{tabular}{@{}c*{15}{c}}
            \hline\hline
            Dep. graph generation rule & $\overline{W}$ & $t_{\mathrm{dep}}\ (\mathrm{s})$ & $t_{\mathrm{sch}}\ (\mathrm{s})$ & $t_{\mathrm{tot}}\ (\mathrm{s})$ & $gap\ (\%)$ \\
            \hline
            General & 1.63 & 133.4 & 386.4 & 519.8 & 0.00 \\
            Trivial & 1.06 & \0\00.3 & 340.5 & 340.8 & 1.91 \\
            Serial & 1.00 & $-$ & $-$ & $-$ & 6.11 \\
            \hline\hline
    \end{tabular}
\end{table}

\begin{table}
    \caption{\label{tab:gap_to_serial_scheduling} Average gap to the LSSP solution generated using the serial rule. As the total required number of qubits $N$ and the average percentage of qubits required per operation $N_{\%}$ increase, solutions tend to become serialized as the average gap converges to 0\%. Therefore, better dependency graph generation rules have no advantage over the serial rule. Conversely, with fewer qubits required per operation, the potential for circuit parallelization increases.}
    \centering
    \begin{tabular}{@{}c*{15}{c}}
        \hline\hline
        Dep. graph generation rule & \textbf{$N$} & \textit{$N_{\%}=0.15$} & \textit{$N_{\%}=0.50$} & \textit{$N_{\%}=0.85$}  \\
        \hline
        \multirow{3}{*}{General}         & 10 & 38.73 & 2.26 & 3.98\\
                                    & 30 & \09.05 & 0.00 & 0.01\\
                                    & 50 & \01.00 & 0.00 & 0.00\\
        \hline
        \multirow{3}{*}{Trivial}    & 10 & 30.35 & 0.84 & 0.00\\
                                    & 30 & \04.34 & 0.00 & 0.00\\
                                    & 50 & \00.16 & 0.00 & 0.00\\
        \hline\hline
    \end{tabular}
\end{table}

The trivial rule is computationally lighter than the general rule, with a shorter average wall-clock time for dependency graph generation and scheduling. Its efficiency increases with the number of rotations, making it ideal for large-scale circuits. Although its solutions are, on average, 1.91\% worse than the best known solutions, there is potential for improvement if simultaneous data qubit contributions to multiple measurements are allowed. Serial solutions are, on average, 6.11\% worse than the best known solutions, but the gap varies with circuit characteristics. For circuits with N = 10 and $N_\%$ = 0.15, parallel solutions can be 30--38\% better than serial ones. However, as more qubits are requested per operation, optimal solutions tend to align with serial scheduling due to reduced parallelization potential.

Based on these observations, we conclude that the trivial rule offers better scalability for circuits with greater length without having a significant impact on the solution quality. In \cref{app:real_circuits_summary}, we show that, even after the transpilation, the Hamiltonian simulation circuits generated can require millions of operations to be scheduled. Consequently, for the remaining experiments, we use the trivial rule as the dependency graph generation rule.

\subsection{Transpilation analysis} \label{sec:gate_opt_analysis}

In \cref{tab:real_circuits_results}, we provide a comparison of solutions for the LSSP on the Hamiltonian simulation circuits using the trivial rule. To capture substantial parallelizability without dramatically increasing space costs, we set $|\mathcal{V}^\mathrm{S}| = |\mathcal{V}^\mathrm{A}| = \left\lceil \,\overline{W} \,\right\rceil$ for these runs. The columns with the headings \textit{Pre-transpiled Circuit} and \textit{Post-transpiled Circuit} present the scheduling results for the circuits before and after the transpilation described in \cref{app:improved_gate_optimization}, respectively.
The following are some key observations from the data in the table.

\begin{table}
    \caption{\label{tab:real_circuits_results}Summary of results for the Hamiltonian simulation circuits. $\mathbb{E}(N)$: the expected number of logical cycles needed to execute the generated schedule; $LB$: the lower bound for $\mathbb{E}(N)$ given by the circuit depth; $U\!B$: the upper bound for $\mathbb{E}(N)$ given by the circuit length; $t$ (s): the total wall-clock time taken to execute the earliest-available-first scheduling algorithm, in seconds.}
    \small
    \begin{tabular*}{\textwidth}{@{}lr*{8}{@{\extracolsep{0pt plus 12pt}}r}}
    \hline\hline
     & \multicolumn{4}{c}{Pre-transpiled Circuit} & \multicolumn{4}{c}{Post-transpiled Circuit}\\
    \vspace{-6mm}\\
    &\crule{4}&\crule{4}\\
    Circuit & \multicolumn{1}{c}{$\mathbb{E}(N)$} &  \multicolumn{1}{c}{$LB$} &  \multicolumn{1}{c}{$U\!B$}& \multicolumn{1}{c}{$t\ (\mathrm{s})$} &\multicolumn{1}{c}{\textit{$\mathbb{E}(N)$}} &\multicolumn{1}{c}{$LB$} &  \multicolumn{1}{c}{$U\!B$} & \multicolumn{1}{c}{$t\ (\mathrm{s})$} \\
    \hline
    H$_2$       & 126,874 & 111,458       & 155,845   & 3.5       & 17,967    & 16,004    & 20,190    & 0.6\\
    LiH       & 5,996,028     & 5,969,511     & 6,165,707 & 235.2     & 499,056   & 498,242   & 499,956   & 16.0 \\
    H$_2$O      & 18,342,657    & 18,308,449    & 18,674,983 & 876.6   & 1,570,803 & 1,568,627 & 1,572,334 & 54.1 \\
    Chain10   & 161,381   & 127,456       & 245,558   & 34.1      & 16,209    & 11,850    & 26,094    & 3.7\\
    Chain20   & 1,480,722     & 1,210,702     & 2,439,547 & 1401.0     & 178,039   & 134,262   & 276,383   & 43.5\\
    Chain30   & 2,289,993     & 1,845,468     & 3,718,281 & 4244.8    & 266,178   & 205,402   & 421,462   & 148.0 \\
    Chain40   & 3,047,568     & 2,494,727     & 5,025,943 & 11764.1    & 362,395   & 276,593   & 568,238	& 184.3 \\
    Chain50   & 3,787,624     & 3,128,340     & 6,286,923 & 20,774.4    & 457,277   & 345,321   & 712,165 & 322.3 \\
    TFSK10   & 1,255,636     & 1,045,161     & 3,171,274	& 103.2     & 356,734   & 322,515   & 405,152	& 9.1 \\
    TFSK15   & 2,235,327     & 1,630,190     & 6,916,378 & 1084.8     & 793,954 & 716,148 & 890,287 & 23.0 \\
    Heisenberg5 & 1,051,046 & 989,580 & 1,451,190 & 35.4 & 161,605 & 145,974	&177,390 &	3.4 \\
    Heisenberg10 & 2,062,007	& 1,876,235	& 2,914,850	& 144.1 & 309,846 & 279,904&	349,330	&7.1 \\
    Heisenberg20 & 4,042,805	& 3,598,315 & 5,790,940 & 1346.1 & 610,760 &	523,780 & 707,142 & 23.2 \\
    Heisenberg30 & 5,980,488 & 5,343,285 & 8,689,920 & 4735.9 & 802,016 &	685,194 &	1,023,910 &	25.4 \\
    Heisenberg40 & 7,838,221 & 7,027,555 & 11,528,200 & 13,629.8 & 1,196,034 &	1,028,524 &	1,394,720 &	56.5 \\
    RSC3 &83,002 &	69,610 & 145,781 & 4.0 & 10,332 & 7800 & 16,175 &	0.7 \\
    RSC5 & 712,734 & 587,120 &	1,961,200 &	86.5 & 143,902 &	75,726 & 216,642 & 31.4\\
    RSC7 & 1,042,185 & 804,273 & 3,707,678 & 566.0 & 255,207 &	130,736 &	396,707 & 77.1\\
    \hline
    Average         & 3,418,683     & 3,120,413     & 4,943,900 & 3392.8    & 444,906   & 387,367   & 537,460   & 57.2\\
    \hline\hline
    \end{tabular*}
\end{table}\normalsize

\begin{itemize}
\item Our proposed algorithm efficiently solves the LSSP for the Hamiltonian simulation circuits within a reasonable length of time. It can schedule approximately 20,000 operations per second, on average, with a 60\% faster performance on post-transpiled circuits compared to pre-transpiled ones.
\item Serial scheduling (column $U\!B$) solutions are improved by 29.5\%, on average, when operations are parallelized. This reduction is more pronounced for pre-transpiled circuits (37.3\%) compared to post-transpiled ones (21.7\%), exemplifying the benefits of parallelization, regardless of transpilation.
\item Across all tested circuits, the lower bound of pre-transpiled circuits exceeds the upper bound of post-transpiled ones significantly. This highlights the effectiveness of transpilation in reducing $\mathbb{E}(N)$, contradicting arguments that reduced parallelizability leads to prohibitive runtimes \cite{beverland2022surface}. On average, our experiments demonstrate an 89\% reduction in circuit length and an 84\% reduction in $\mathbb{E}(N)$ after transpilation.
\end{itemize}

\section{Conclusion}
\label{sec:conclusion}

Our study has investigated the lattice surgery scheduling problem (LSSP), which determines the sequencing of lattice surgery operations on a two-dimensional architecture consisting of topological error correcting codes. A logical layout of the architecture guides resource allocation for quantum operations. Operations requiring multiple qubits require the creation of ancilla patches for the entanglement of the required qubits. We optimize ancilla patches by solving terminal Steiner tree problems to minimize the execution time. Parallel scheduling is investigated by involving ancilla patches representing a quantum bus surrounding data qubits within a dedicated central zone on the layout.

We decompose the LSSP into subproblems based on a forest packing problem. Since enumerating all sets of candidate operations for parallelization is impractical, an algorithm based on the earliest-available-first policy is implemented to select candidates for parallelization, after which a greedy algorithm is used to solve the forest packing problem for the selected operations, taking layout constraints into account.

Our computational experiments reveal that employing a trivial rule to generate dependency constraints enhances scalability for larger circuits. Application-inspired large-scale circuits, comprising up to 18 million quantum gates and 50 qubits, are successfully scheduled within reasonable time frames. We show that parallel scheduling reduces the expected circuit execution time, but it is heavily dependent on the structure of the logical circuit being scheduled. Also, LSSP solutions for optimized circuits outperform scheduling for non-transpiled circuits, reducing the expected number of logical cycles needed to execute the generated schedule by around one order of magnitude in all circuits tested.

There exist cases where transpilation can increase circuit depth, such as in circuits with sequential CNOT gates acting on different qubits followed by sequences of commuting $T$ gates, which would make them lose commutativity after transpilation. However, in general, the removal of Clifford gates is expected to be highly beneficial in reducing FTQC execution time. While our approach considers the scheduling of operations in the Clifford + $T$ gate set, there exist architectures that might improve runtimes using operations in a different basis, such as when arbitrary-angle gates are left in the circuit to be synthesized using resources generated externally to the central zone. In such cases, partial transpilation can still be used to reduce circuit depth in parts of the circuit involving Clifford + $T$ gates.

Finally, the proposed layout considers $Y$ operators to be common in the transpiled circuit. In cases where this is not observed, using more-compact layouts with a smart placement of qubits may be desirable to reduce space costs.  A future research direction could be to perform a similar scheduling analysis as was performed in this paper on a colour code lattice, as colour codes allow for easy access to each Pauli measurement, and may lead to more-compact layouts. Our study is a contribution to research on the development of scalable quantum compilers and provides valuable insights into estimating quantum resources required for future fault-tolerant computations.

\appendix

\section{Efficient transpilation} \label{app:improved_gate_optimization}

The transpilation of Clifford operations out of the circuit works by
transforming all of the gates in a circuit into Pauli rotations, and then
commuting the Clifford operations past each $\pi/8$ rotation arising from a $T$
gate \cite{litinski2018game}. The main difficulty with this approach is that
the set of rotation gates is not closed under multiplication, even when the
rotation gates are restricted to Clifford operations, and since we need to
preserve the order of commutation this requires an algorithm reminiscent of
bubble sort. Hence, the runtime is $\mathcal{O}(m^2)$, where $m$ is the length
of the circuit, since we need to push each Clifford gate past each $\pi/8$
rotation gate individually.

This $\mathcal{O}(m^2)$ algorithm can be avoided, however, if we use the
symplectic representation of Clifford gates \cite{kim2022fault-tolerant}. Since Clifford gates can be
combined, this allows us to perform operations corresponding to multiple
rotation commutations in a single step. We perform a simple pass through the
circuit, keeping an accumulated Clifford operation representing all of the
Clifford operations seen thus far, and at each step either commute this
accumulator through a $\pi/8$ gate or combine it with the Clifford gate,
depending on the gate present at that particular step. This reduces the runtime
of the main optimization step from $\mathcal{O}(m^2)$ to $\mathcal{O}
(m)$, leading to massive reductions in computational costs.

\subsection{Commuting Clifford and rotation gates}

It is important to understand how commuting a Clifford rotation past a $\pi/8$ rotation
affects it. Let us assume that $U$ is an $n$-qubit unitary and $P$ is an
$n$-qubit Pauli operator. Given an arbitrary angle $\theta$, we can then expand
the conjugation of the Pauli rotation about $P$ by $U$ as
\begin{align}
    U \exp( i \theta P) &= U \exp( i \theta P) U^\dag U = U \left[ \cos(\theta)\mathbb{I} + i \sin(\theta) P \right] U^\dag U\nonumber\\
        &= \left[ \cos(\theta) U \mathbb{I} U^\dag + i \sin(\theta) U P U^\dag\right] U = \exp( i \theta UPU^\dag) U.
\end{align}
Hence, if $U$ is a Clifford operation such that $UPU^\dag$ is also a Pauli matrix (a defining feature of Clifford operations), commuting $U$ past a Pauli rotation results in $U$ remaining unchanged, while the Pauli operator is updated through conjugation. Since this works for arbitrary angles, it
also works in the case of $\pi/8$ rotations.

\subsection{Representing Clifford operations}

The tableau representation of Clifford gates has been extremely useful in
simulating stabilizer circuits and states~\cite{gidney2021stim}, and is extremely useful in optimizing transpilation~\cite{kim2022fault-tolerant}. We leverage the ease of representing multiplication of Clifford
operators and the conjugation of a given Pauli matrix by Clifford operators. We
also note that it is straightforward to transform a given $\pi/4$ rotation into
such a representation.
Conjugation can be understood as a relatively straightforward combination of rows of the tableau, along with some bookkeeping to keep
track of the phase, and multiplication as a sequence of conjugation calls. Further, the initialization of $\pi/4$ rotations can be instantiated through multiplication of Pauli matrices.

For a given $n$-qubit Pauli operator $P$, its symplectic representation is a list of $2n+1$ bits, where the first bit corresponds to the phase of the Pauli operator, the next $n$ bits represent the $X$ generators of the Pauli matrix, and the last $n$ bits represent the $Z$ generators of the Pauli matrix. In particular, we say that $P = (\theta|x|z)$, where $x$ and $z$ are
each $n$-bit strings. We can then determine the Pauli element to which $P$ corresponds by inspecting the values of $x$ and $z$.
For a given qubit $j$, $P$ acts on qubit $j$ as: the identity matrix if $x_j$ and $z_j$ are equal to 0; $X$ if $x_j = 1$ and $z_j = 0$; $Y$ if
$x_j = 1$ and $z_j = 1$; and $Z$ if $x_j = 0$ and $z_j = 1$. Note that, since $i XZ = Y$, if both $x_j$ and $z_j$
are equal to 1, we have acquired an additional factor of $i$ that is not accounted for elsewhere in the representation.

\subsection{An improved transpilation algorithm}

We keep track of the
image of a generating set of the Pauli group under conjugation by the Clifford operator $C$. While the specific order of the generators does not
matter, in our representation we alternate between single-qubit $X$ and $Z$ operations acting on qubits of increasing indices. As an example,
\begin{align*}
    C &= \left( \begin{array}{c|cc|cc}
        0 & 1 & 1 & 0 & 0\\
        0 & 0 & 0 & 1 & 0\\
        0 & 0 & 1 & 0 & 0\\
        0 & 0 & 0 & 1 & 1
    \end{array}\right)
\end{align*}
represents a CNOT operation from the first qubit to the second. In this representation, an $X$ generator on the first
qubit is mapped to an $XX$ generator when conjugated by this Clifford operator.
The improved transpilation algorithm is shown in \cref{alg:improved_circuit_optimization}. We
simply sweep through the circuit, either updating the given gate
through conjugation or updating the accumulated Clifford operator, resulting in
a runtime that grows linearly with the length of the circuit. In this
pseudo-code, for a given rotation $R$, $R_\theta$ denotes the angle of the
rotation and $R_{\text{Pauli}}$ denotes the Pauli operator of the rotation. We
now describe the various update procedures needed on the Clifford tableau
representations to implement this algorithm.

\begin{algorithm}[b]
   \footnotesize
   \caption{~Efficient Transpilation Algorithm}
   \label{alg:improved_circuit_optimization}
   \begin{algorithmic}[1]
   \STATE \textbf{input} Pauli rotation circuit ($\mathcal{R}$);
    \STATE Let $C= $ tableau$(\mathbb{I})$;
    \STATE Let $\mathcal{R}'$ be an empty set of rotations;
    \FOR {$R \in \mathcal{R}$}
        \IF {$R_\theta = \pi/8$}
            \STATE Let $R_{\text{Pauli}} = C.$conjugate$(R_{\text{Pauli}})$;
            \STATE Append $R$ to $\mathcal{R}'$;
        \ELSIF {$R_\theta = \pi/4$ or $\pi/2$}
            \STATE Let $C' = $ tableau$(R)$;
            \STATE Let $C = C \times C'$;
        \ELSIF {$r$ is a measurement}
            \STATE Let $R_{\text{Pauli}} = C.$conjugate$(R_{\text{Pauli}})$;
            \STATE Append $R$ to $\mathcal{R}'$;
        \ENDIF
    \ENDFOR
    \RETURN $\mathcal{R}'$
   \end{algorithmic}
\end{algorithm}

\paragraph*{Conjugation.}

With this representation, we can determine the action of conjugation by this Clifford operator on an arbitrary Pauli operator by analyzing
the action of each generator making up the given Pauli operator. In particular, if the Pauli operator is given by the symplectic representation
$P = (\theta | x | z)$, where $x$ and $z$ are bit strings of length $n$ and $\theta$ is a single bit, then there are two main computations
we need to determine: to which Pauli operator the operation gets mapped and the overall phase to which to map.

To determine the particular Pauli operator to which $P$ gets mapped under the Clifford operator $C$, we look at the bits of $x$ and $z$ that are nonzero and perform an \textsc{XOR} operation between the rows of $C$ that correspond to nonzero bits. If we represent the row of $C$ corresponding
to a Pauli operator $P$ as $C_P$, and if $C_P^b$ refers to the all-zeroes string if $b$ is zero or refers to $C_P$ if $b$ is one, we can define the
Pauli operators $P_x = C_{X_1}^{x_1} \oplus C_{X_2}^{x_2} \oplus \cdots \oplus C_{X_n}^{x_n} = (\theta_x| x_x |z_x)$ and $P_z = C_{Z_1}^{z_1} \oplus C_{Z_2}^{z_2}
\oplus \cdots \oplus C_{Z_n}^{z_n} = (\theta_z | x_z | z_z)$.  Then, the Pauli operator to which $P$ is mapped is $P_x \oplus P_z$.

After determining the basis of the given Pauli operator, we must still determine its phase under the mapping. One necessary component is determining the number of
commutations that occur when combining each of the $C_{X_i}$ and $C_{Z_i}$, which can be found by iteratively calculating the $z$ operator for increasing $i$, and taking the inner product with the $x$ operator of the $C_{X_i}$.

Another feature that affects the final phase is the result of the factors of $i$. The number of $i$'s initially in $P$, the number of $i$'s created when mapping to $P_x$ and $P_z$, and the number of $i$'s in the representation of $P_x \oplus P_z$ combine to define the final phase of the operator. Specifically, the number of initial $i$'s
is given by $n_{i,i} = |x \cdot z|$, the number of intermediate $i$'s is $n_{i,m} = \sum_j |C_{X_j,x} \cdot C_{X_j,z}| + |C_{Z_j,x} \cdot C_{Zj,z}|$, and the number of final $i$'s is given by $n_{i,f} = |(x_x \oplus x_z) \cdot (z_x \oplus z_z)|$.  The final change in the phase resulting from the number of $i$'s is then given by $ (n_{i,i} + n_{i,m} - n_{i,f})/2 \mod 2 = \theta_i$.
Putting all of the above together, the mapping of $P$ under the conjugation of $C$ is given by $CPC^\dag = (\theta \oplus
 \theta_x \oplus \theta_z \oplus \theta_c \oplus \theta_i | x_x \oplus x_z | z_x \oplus z_z)$.

\paragraph*{Multiplication.}

Another attribute needed to implement the improved transpilation algorithm is the ability to multiply Clifford operations. In
particular, if $U$ and $V$ are Clifford operations, the mapping of $UV P V^\dag U^\dag$ for each
generator of the Pauli group must be determined. Fortunately, we already have the mapping $V P V^\dag$ from the representation of $V$.  To determine
the full mapping, we need to determine the conjugation of this operation by $U$, for which we can use the previous algorithm
for conjugation. A given row of $UV$ is then determined by the conjugation of the corresponding row of $V$ by $U$.

\paragraph*{Initialization.}
The attribute needed is the ability to initialize a Clifford representation from our representation of Pauli rotations, that is, how given $\pi/4$ and $\pi/2$ rotations affect each Pauli generator under conjugation.
To understand how a $\pi/2$ rotation about a Pauli operator $P$ affects another Pauli operator $Q$ under conjugation requires an explicit calculation of the commutation according to the equation
 \begin{align}
    \exp( i \pi /4 P) Q \exp(-i \pi/4 P) &= \left( \cos(\pi/4) \mathbb{I} + i\sin(\pi/4) P\right) Q \left( \cos(\pi/4) \mathbb{I} - i\sin(\pi/4) P\right) \nonumber\\
      &= \frac{1}{2} \left(Q + i P Q - i Q P + PQP\right).
 \end{align}
 From this expression, we can deduce that if $P$ and $Q$ commute, the $\pi/4$ rotation maps $Q$ to itself. Similarly, if $P$ and $Q$ do not commute, then this rotation maps to $i P Q$.

\section{Characteristics of the Hamiltonian simulation circuits} \label{app:real_circuits_summary}

\cref{tab:real_circuits_summary} summarizes the key characteristics of the Hamiltonian simulation circuits generated, as described in \cref{sec:test_circuits}, before and after the transpilation described in \cref{app:improved_gate_optimization}.

\begin{table}[H]
    \caption{\label{tab:real_circuits_summary}Summary of the Hamiltonian simulation circuits. ``Circuit'': the name of the circuit; $N$: the total number of qubits required by the circuit; $|\mathcal{R}|$: the circuit length (the number of operations in the circuit); $|\mathcal{R}^{\pi/4}|$ and $|\mathcal{R}^{\pi/8}|$: the number of $\pi/4$ and $\pi/8$ rotations in the circuit, respectively; $\overline{W}$: the average dependency graph width (\cref{sec:scalability_analysis}); and $N_{\%}$: the average number of qubits per operation.}
    \small
    \begin{tabular*}{\textwidth}{@{}lcr*{7}{@{\extracolsep{0pt plus 12pt}}r}}
        \hline\hline
         & & \multicolumn{5}{c}{Pre-transpiled Circuit} & \multicolumn{3}{c}{Post-transpiled Circuit}\\
        \vspace{-6mm}\\
        &&\crule{5}&\crule{3}\\
        Circuit & $N$ & \multicolumn{1}{c}{$|\mathcal{R}|$} & \multicolumn{1}{c}{$|\mathcal{R}^{\pi/4}|$} & \multicolumn{1}{c}{$|\mathcal{R}^{\pi/8}|$} & \multicolumn{1}{c}{$\overline{W}$} & \multicolumn{1}{c}{$N_{\%}$} & \multicolumn{1}{c}{$|\mathcal{R}|$} & \multicolumn{1}{c}{$\overline{W}$} & \multicolumn{1}{c}{$N_{\%}$} \\
        \hline
        H$_2$ & 4       & 155,845   & 103,441 & 52,404 & 1.40 & 1.00    & 20,190    & 1.26 & 2.31\\
        LiH & 12      & 6,165,707 & 3,394,011 & 2,771,696 & 1.03 & 1.00 & 499,956   & 1.00 & 8.68\\
        H$_2$O & 14     & 18,674,983 & 9,750,953 & 8,924,030 & 1.02 & 1.00  & 1,572,334 & 1.00 & 10.35\\
        Chain10 & 10  & 245,558   & 155,418 & 90,140 & 1.93 & 1.00  & 26,094    & 2.20 & 1.89\\
        Chain20 & 20  & 2,439,547 & 1,522,062 & 917,485 & 2.01 & 1.00   & 276,383   & 2.06 & 2.31\\
        Chain30 & 30  & 3,718,281 & 2,334,375 & 1,383,906 & 2.01 & 1.00 & 421,462   & 2.05 & 2.52\\
        Chain40 & 40  & 5,025,943 & 3,137,085 & 1,888,858 & 2.01 & 1.00  & 568,238   & 2.05 & 2.14\\
        Chain50 & 50  & 6,286,923 & 3,920,154 & 2,366,769 & 2.01 & 1.00 & 712,165 & 2.06 & 2.18\\
        TFSK10 & 10  & 3,171,274 & 2,204,520 & 966,754 & 3.03 & 1.00 & 405,152   & 1.26 & 3.78\\
        TFSK15 & 15  & 6,916,378 & 4,806,264 & 2,110,114 & 4.24 & 1.00  & 890,287   & 1.24 & 4.89\\
        Heisenberg5 & 5 & 1,451,190 & 959,415 & 491,775 &	1.47 & 1.00 & 177,390 & 1.22 & 2.82 \\
        Heisenberg10 & 10 & 2,914,850 &1,935,570 & 979,280 &  1.55 & 1.00 & 349,330 & 1.25 & 2.90\\
        Heisenberg20 & 20 & 5,790,940 & 3,818,520 & 1,972,420 & 1.61 & 1.00 & 707,142 & 1.35 & 2.57 \\
        Heisenberg30 & 30 & 8,689,920 & 5,735,250 & 2,954,670 & 1.63 & 1.00 & 1,023,910 & 1.49 & 1.84\\
        Heisenberg40 & 40 & 11,528,200 & 7,563,480 & 3,964,720 & 1.64 & 1.00 & 1,394,720 & 1.36 & 2.49\\
        RSC3 & 9 & 145,781 & 88,299 & 57,482 & 2.09 &1.00 & 16,175 & 2.07 & 1.98\\
        RSC5 & 25 & 1,961,200 & 1,167,711 & 793,489 & 3.34 &	1.00 & 216,642 & 2.86 & 2.63\\
        RSC7 & 49 & 3,707,678 &2,166,480 & 1,541,198 & 4.61& 1.00 & 396,707 & 3.03& 3.15\\
        \hline\hline
    \end{tabular*}
\end{table}\normalsize

\bibliography{references}

\begin{thebibliography}{10}

\bibitem{arrazola2023differentiable}
J.~M. Arrazola, S.~Jahangiri, A.~Delgado, J.~Ceroni, J.~Izaac, A.~Száva, U.~Azad, R.~A. Lang, Z.~Niu, O.~Di Matteo, R.~Moyard, J.~Soni, M.~Schuld, R.~A. Vargas-Hernández, T.~Tamayo-Mendoza, C.~Y.-Y. Lin, A.~Aspuru-Guzik, and N.~Killoran.
\newblock Differentiable quantum computational chemistry with {P}ennylane, 2023.
\newblock \href {https://arxiv.org/abs/2111.09967} {\path{arXiv:2111.09967}}.

\bibitem{beverland2022surface}
M.~Beverland, V.~Kliuchnikov, and E.~Schoute.
\newblock Surface code compilation via edge-disjoint paths.
\newblock {\em PRX Quantum}, 3(2):020342, 2022.
\newblock \href {https://doi.org/10.48550/arXiv.2110.11493} {\path{doi:10.48550/arXiv.2110.11493}}.

\bibitem{braunstein2018cavity}
A.~Braunstein and A.~P. Muntoni.
\newblock The cavity approach for {S}teiner trees packing problems.
\newblock {\em Journal of Statistical Mechanics: Theory and Experiment}, 2018(12):123401, 2018.
\newblock \href {https://doi.org/10.48550/arXiv.1712.07041} {\path{doi:10.48550/arXiv.1712.07041}}.

\bibitem{bravyi2012magic}
S.~Bravyi and J.~Haah.
\newblock Magic-state distillation with low overhead.
\newblock {\em Phys. Rev. A}, 86(5):052329, 2012.
\newblock \href {https://doi.org/10.1103/PhysRevA.86.052329} {\path{doi:10.1103/PhysRevA.86.052329}}.

\bibitem{bravyi2005universal}
S.~Bravyi and A.~Kitaev.
\newblock Universal quantum computation with ideal {C}lifford gates and noisy ancillas.
\newblock {\em Phys. Rev. A}, 71(2):022316, 2005.
\newblock \href {https://doi.org/10.1103/PhysRevA.71.022316} {\path{doi:10.1103/PhysRevA.71.022316}}.

\bibitem{drake2004approximation}
D.~E. Drake and S.~Hougardy.
\newblock On approximation algorithms for the terminal {S}teiner tree problem.
\newblock {\em Information Processing Letters}, 89(1):15--18, 2004.
\newblock \href {https://doi.org/10.1016/j.ipl.2003.09.014} {\path{doi:10.1016/j.ipl.2003.09.014}}.

\bibitem{fischetti2017thinning}
M.~Fischetti, M.~Leitner, I.~Ljubi{\'c}, M.~Luipersbeck, M.~Monaci, M.~Resch, D.~Salvagnin, and M.~Sinnl.
\newblock Thinning out {S}teiner trees: a node-based model for uniform edge costs.
\newblock {\em Mathematical Programming Computation}, 9(2):203--229, 2017.
\newblock \href {https://doi.org/10.1007/s12532-016-0111-0} {\path{doi:10.1007/s12532-016-0111-0}}.

\bibitem{fowler2012surface}
A.~G. Fowler, M.~Mariantoni, J.~M. Martinis, and A.~N. Cleland.
\newblock Surface codes: Towards practical large-scale quantum computation.
\newblock {\em Phys. Rev. A}, 86(3):032324, 2012.
\newblock \href {https://doi.org/10.1103/PhysRevA.86.032324} {\path{doi:10.1103/PhysRevA.86.032324}}.

\bibitem{gassner2010steiner}
E.~Gassner.
\newblock The {S}teiner forest problem revisited.
\newblock {\em Journal of Discrete Algorithms}, 8(2):154--163, 2010.
\newblock \href {https://doi.org/10.1016/j.jda.2009.05.002} {\path{doi:10.1016/j.jda.2009.05.002}}.

\bibitem{gidney2021stim}
C.~Gidney.
\newblock Stim: a fast stabilizer circuit simulator.
\newblock {\em {Quantum}}, 5:497, 2021.
\newblock \href {https://doi.org/10.22331/q-2021-07-06-497} {\path{doi:10.22331/q-2021-07-06-497}}.

\bibitem{hoang2012steiner}
N.-D. Ho{\`a}ng and T.~Koch.
\newblock Steiner tree packing revisited.
\newblock {\em Mathematical Methods of Operations Research}, 76:95--123, 2012.
\newblock \href {https://doi.org/10.1007/s00186-012-0391-8} {\path{doi:10.1007/s00186-012-0391-8}}.

\bibitem{horsman2012surface}
D.~Horsman, A.~G. Fowler, S.~Devitt, and R.~V. Meter.
\newblock Surface code quantum computing by lattice surgery.
\newblock {\em New Journal of Physics}, 14(12):123011, 2012.
\newblock \href {https://doi.org/10.1088/1367-2630/14/12/123011} {\path{doi:10.1088/1367-2630/14/12/123011}}.

\bibitem{kim2022fault-tolerant}
I.~H. Kim, Y.-H. Liu, S.~Pallister, W.~Pol, S.~Roberts, and E.~Lee.
\newblock Fault-tolerant resource estimate for quantum chemical simulations: Case study on {L}i-ion battery electrolyte molecules.
\newblock {\em Phys. Rev. Res.}, 4(2):023019, 2022.
\newblock \href {https://doi.org/10.1103/PhysRevResearch.4.023019} {\path{doi:10.1103/PhysRevResearch.4.023019}}.

\bibitem{lao2018mapping}
L.~Lao, B.~Van~Wee, I.~Ashraf, J.~Van~Someren, N.~Khammassi, K.~Bertels, and C.~G. Almudever.
\newblock Mapping of lattice surgery-based quantum circuits on surface code architectures.
\newblock {\em Quantum Science and Technology}, 4(1):015005, 2018.
\newblock \href {https://doi.org/10.1088/2058-9565/aadd1a} {\path{doi:10.1088/2058-9565/aadd1a}}.

\bibitem{lau2005packing}
L.~C. Lau.
\newblock Packing {S}teiner forests.
\newblock In {\em International Conference on Integer Programming and Combinatorial Optimization}, pages 362--376. Springer, 2005.
\newblock \href {https://doi.org/10.1007/11496915_27} {\path{doi:10.1007/11496915_27}}.

\bibitem{lin2002terminal}
G.~Lin and G.~Xue.
\newblock On the terminal {S}teiner tree problem.
\newblock {\em Information Processing Letters}, 84(2):103--107, 2002.
\newblock \href {https://doi.org/10.1016/S0020-0190(02)00227-2} {\path{doi:10.1016/S0020-0190(02)00227-2}}.

\bibitem{litinski2018game}
D.~Litinski.
\newblock A game of surface codes: large-scale quantum computing with lattice surgery.
\newblock {\em Quantum}, 3:128, 2019.
\newblock \href {https://doi.org/10.22331/q-2019-03-05-128} {\path{doi:10.22331/q-2019-03-05-128}}.

\bibitem{litinski2019magic}
D.~Litinski.
\newblock Magic state distillation: not as costly as you think.
\newblock {\em Quantum}, 3:205, 2019.
\newblock \href {https://doi.org/10.22331/q-2019-12-02-205} {\path{doi:10.22331/q-2019-12-02-205}}.

\bibitem{ljubic2021solving}
I.~Ljubi{\'c}.
\newblock Solving {S}teiner trees: recent advances, challenges, and perspectives.
\newblock {\em Networks}, 77(2):177--204, 2021.
\newblock \href {https://doi.org/10.1002/net.22005} {\path{doi:10.1002/net.22005}}.

\bibitem{mehlhorn1988faster}
K.~Mehlhorn.
\newblock A faster approximation algorithm for the {S}teiner problem in graphs.
\newblock {\em Information Processing Letters}, 27(3):125--128, 1988.
\newblock \href {https://doi.org/10.1016/0020-0190(88)90066-X} {\path{doi:10.1016/0020-0190(88)90066-X}}.

\bibitem{pajor2018robust}
T.~Pajor, E.~Uchoa, and R.~F. Werneck.
\newblock A robust and scalable algorithm for the {S}teiner problem in graphs.
\newblock {\em Mathematical Programming Computation}, 10:69--118, 2018.
\newblock \href {https://doi.org/10.1007/s12532-017-0123-4} {\path{doi:10.1007/s12532-017-0123-4}}.

\bibitem{ribeiro2002hybrid}
C.~C. Ribeiro, E.~Uchoa, and R.~F. Werneck.
\newblock A hybrid {GRASP} with perturbations for the {S}teiner problem in graphs.
\newblock {\em INFORMS Journal on Computing}, 14(3):228--246, 2002.
\newblock \href {https://doi.org/10.1287/ijoc.14.3.228.116} {\path{doi:10.1287/ijoc.14.3.228.116}}.

\bibitem{senicourt2022tangelo}
V.~Senicourt, J.~Brown, A.~Fleury, R.~Day, E.~Lloyd, M.~P. Coons, K.~Bieniasz, L.~Huntington, A.~J. Garza, S.~Matsuura, R.~Plesch, T.~Yamazaki, and A.~Zaribafiyan.
\newblock Tangelo: An open-source {P}ython package for end-to-end chemistry workflows on quantum computers, 2022.
\newblock \href {https://arxiv.org/abs/2206.12424} {\path{arXiv:2206.12424}}.

\bibitem{sun2022packing}
Y.~Sun, G.~Gutin, and X.~Zhang.
\newblock Packing strong subgraph in digraphs.
\newblock {\em Discrete Optimization}, 46:100745, 2022.
\newblock \href {https://doi.org/10.1016/j.disopt.2022.100745} {\path{doi:10.1016/j.disopt.2022.100745}}.

\bibitem{uchoa2012fast}
E.~Uchoa and R.~F. Werneck.
\newblock Fast local search for the {S}teiner problem in graphs.
\newblock {\em ACM Journal of Experimental Algorithmics}, 17:1--22, 2012.
\newblock \href {https://doi.org/10.1145/2133803.2184448} {\path{doi:10.1145/2133803.2184448}}.

\bibitem{watkins2024high}
G.~Watkins, H.~M. Nguyen, K.~Watkins, S.~Pearce, H.-K. Lau, and A.~Paler.
\newblock A high performance compiler for very large scale surface code computations.
\newblock {\em {Quantum}}, 8:1354, 2024.
\newblock \href {https://doi.org/10.22331/q-2024-05-22-1354} {\path{doi:10.22331/q-2024-05-22-1354}}.

\end{thebibliography}

\end{document}